# High-Throughput Identification of Electrides from all Known Inorganic Materials.

Lee A. Burton, Francesco Ricci, Gian-Marco Rignanese, Geoffroy Hautier. *

Institute of Condensed Matter and Nanoscience, Université catholique de Louvain, 1348 Louvain-la-Neuve, Belgium

**ABSTRACT:** In this paper, we present the results of a large-scale, high-throughput computational search for electrides among all known inorganic materials. Analyzing a database of density functional theory results on more than 60,000 compounds, we identify 69 new electride candidates. We report on all these candidates and discuss the structural and chemical factors leading to electride formation. Among these candidates, our work identifies the first partially-filled 3d transition metal containing electrides $Ba_3CrN_3$ and $Sr_3CrN_3$; an unexpected finding that contravenes conventional chemistry.

Electrides are rare ionic compounds in which an electron does not occupy an atomic orbital but rather acts as an anion. Such an electron is expected to behave differently to those occupying the valence state of standard materials, making electrides desirable as electron emitters,[1] nonlinear optical switches,[2] superconductors,[3] battery anodes,[4] and catalysts for applications ranging from compound synthesis to $CO_2$ splitting.[5,6]

Chemically independent electrons were first reported from salts of metal-ammonia solutions.[7] Solid-state organic electrides were reported in 1990 but were unstable at room temperature even in inert atmosphere.[8] In 2012, a breakthrough came with the first inorganic electride stable at room temperature and in air: $Ca_{12}Al_{14}O_{32}$ in the mayenite structure.[9] Now, various stable electrides are known. They are often classified according to the localization of the anionic electron in the lattice: 0-dimensional (0D)[10] when it resides into a cavity, 1-dimensional (1D)[11] when it is enclosed in a channel, and 2-dimensional (2D)[12] when it is confined to a layer. These compounds are now being tested in multiple industrial applications from $NH_3$ catalysis,[13] to organic light emitting devices (OLED).[14] As such, electride chemistry has become an active field of solid state chemistry and materials science with large efforts directed towards their discovery, synthesis and characterization.

Several simple rules have been used to identify possible electride candidates. Firstly, the structure must contain free space where the anionic electron can reside,[8] be this a cavity, a channel or a layer. Secondly, a compound must have an excess electron to contribute to the lattice, as determined by the sum of oxidation states of the components.[15] Finally, the compound must contain a strongly donating cation to off-set the naturally large energy of a unbound electron.[15] For example, the 2D electride $Ca_2N$ has two $Ca^{2+}$ cations and one $N^{3-}$ anion (according to commonly held concepts of oxidation state). The electropositive Ca cations wish to donate four electrons but the anion can only accept three, leading to an excess electron occupying free space in the crystal structure; thereby satisfying all three criteria. The need for these three factors to combine adequately makes electrides a rare occurrence among solid compounds with only a handful electrides known so far.

Quantum-mechanical calculations have been essential in the quest for new electrides. They are commonly used in conjunction with experimental characterization (e.g., catalytic activity or transport measurements) to demonstrate the electride nature of a material. By observing the localization of electrons resulting from computations in the density functional theory (DFT) framework, one can determine if the electrons occupy space away from a nuclei, the definition of an electride.[16,17] *Ab initio* techniques have, however, been mainly used so far *a posteriori* to rationalize experimental results and only in rare occasion to predict new electrides.[18–20]

Here, we propose an entirely new approach. Using the large database of DFT computed electronic structures for known, stable materials available from the Materials Project, originating mainly from the Inorganic Crystal Structure Database (ICSD),[21] we assess more than sixty thousand compounds and identify 69 new potential electrides. We discuss the three most popular groups of candidate based on chemical structure, which incidentally form 0D, 1D and 2D electrides. Furthermore, we highlight what we believe to be the only known instances of electrides containing a redox active element: $Ba_3CrN_3$ and $Sr_3CrN_3$ (significant as one would expect a redox active transition metal to accept the anionic electron). Finally, we reevaluate the previously proposed guiding principles for electride formation in light of our large-scale screening.

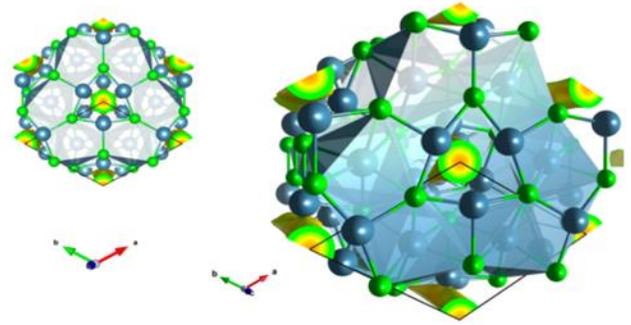

Figure 1: Structure and electron density channel (perpendicular to the page) for the $Mn_5Si_3$ compounds.

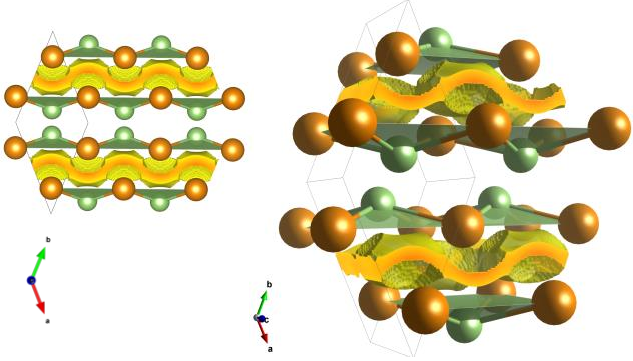

Figure 2: Electron density around $E_F$ for the CrB class candidates clearly indicative of a 2D electride material.

Our computational screening began with the Materials Project,[21] a database of 69,640 materials computed using the Vienna *ab initio* Simulation Package (VASP).[22,23] We focused on materials with the Fermi level $E_F$ crossing several bands *i.e.* metals. The first screening step was using the projections on atomic orbitals for the eigenstates around $E_F$ in a range of 0.01 eV. If these projections summed to less than 50% of the total electrons present, we expect a high chance for the electrons to occupy non-atomic sites and thus, be an electride. In a second step, we performed a Bader charge analysis to identify electron distributions with a relative charge of at least 0.1 located 2.2 Å or more away from the nearest atom,[24–26] within 0.2 eV of the $E_F$ for a 2,000 $k$ points per Å$^{-3}$ resolution calculation. This screening returns 69 non-elemental candidates including the four known electrides $Ca_2N$,[27] $Y_2C$,[28] $Y_5Si_3$,[16] and $LaH_2$.[29] We note that only "as-is" electrides will be found through our approach and that we do not expect to find mayenite for instance, which requires a chemical treatment (oxygen deintercalation) to produce its electride form.[30]

The most common structure among the top electrides is found for equiatomic binary species. These are PrGa, CaAu, NdGa, CaSi, SrSn, BaGe, SrGe, CaGe, BaSi, SrSi and BaSn, which are orthorhombic with the space group Cmcm (N° 63). The structures consist of bonded atomic bilayers sandwiched between 2D free electron layers (see Figure 1) and are referred to as CrB-type structures in the original reports. Of these 11, 8 are II-IV compounds, complying with the principle of including an electropositive element (group II) in the compound. However, the rule of excess electron from oxidation states does not apply. To the knowledge of the authors, these compounds have never been considered electrides but show similar 2D behaviour to the well-studied $Ca_2N$.

Of the 69 systems, 9 adopt the so-called $Mn_5Si_3$ structure type, including the known electride $Y_5Si_3$, which is reported to be chemically stable even in the presence of moisture.[16] These are $Ba_5As_3$, $Ca_5Sb_3$, $Nd_5Ge_3$, $Sr_5As_3$, $Y_5Si_3$, $Sr_5Sb_3$, $Sr_5Bi_3$, $Ba_5Bi_3$ and $Ba_5Sb_3$, all of which are hexagonal with the space group $P6_3/mcm$ (N° 193). Of the 9 compounds, 7 are II-V compounds. In the unit cells, 3 of 5 cations sit at the center of face-sharing octahedra, while the remaining two coordinate towards a 1D channel populated by the anionic electron; see Figure 2.

8 of the systems are tetragonal $X_5Y_3$ compounds with the space group I4/mcm (N° 140). These are $La_5Si_3$, $Ba_5Sn_3$, $Sr_5Sn_3$, $Sr_5Ge_3$, $Ca_5Ge_3$, $Ca_5Au_3$, $Sr_5Si_3$, $Ca_5Si_3$. Most of these systems (6/8) are II-IV compounds, as are those of the CrB type, the exceptions being, in both cases, either Au or f-block elements. Their structure consists of square planar and 5-coordinated cations in distorted square-based pyramid polyhedral – where the base of the pyramid orientates towards a cavity containing localised electron distribution, see Figure 3. The anionic electron completes an octahedral coordination environment for the cation; itself being 4-coordinated.

The previous discussion has focused on the 3 most common groups by structure type in our results. Subsequently, we highlight two compounds identified by our method that would be the first electrides to contain redox active elements: $Ba_3CrN_3$ and $Sr_3CrN_3$. These nitrides exhibit structures composed of trigonal planar Cr and pseudo-tetrahedral coordinated cations (Ba/Sr) in a hexagonal system with the space group $P6_3/m$ (N° 176). In this case, each Ba/Sr is exposed to a 1D anionic electron, similar to the $Mn_5Si_3$ compounds.

So far, the Cr compounds have not been considered as electrides though bond valence calculations indicate an overall excess electron for the compound based on obtained charge of 4.17 and 4.11 for Cr in $Sr_3CrN_3$ and $Ba_3CrN_3$ respectively (assuming $Sr/Ba^{2+}$ and $N^{3-}$).[31] While the authors note that "the values deviate significantly from the expected oxidation states" they ascribe such deviations to bond convalency. They also note an apparent absence of Jahn-Teller distortion for the Cr coordination in this case, which in other similar cases was induced by low-spin $d^3$ $Cr^{III}$, but attribute this to a

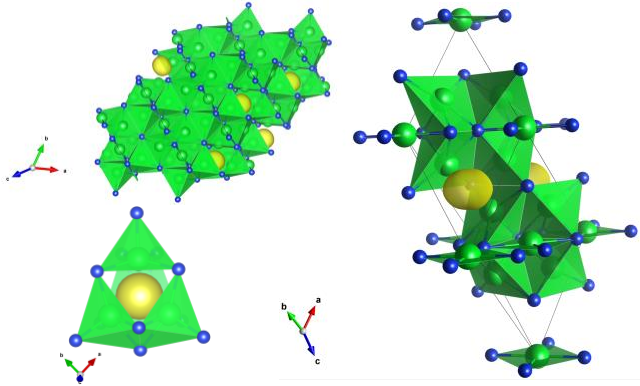

*Figure 3 (left) the supercell structure of $Sr_5Si_3$ (Sr as green, Si as blue); (right) the unit cell and (bottom left) the immediate bonding environment of the 0D free electron density, yellow.*

difference in size for the alkaline-earth cations in these cases. Our results indicate an alternative explanation where $Cr^{IV}$ is formed with an anionic electron in the channel. This finding is counter-intuitive as it could be expected that the energy induced by changing the redox active Cr from 3+ to 4+ would always be more favorable than delocalizing an electron in the channel, as has been reported explicitly for high-pressure electrides.[32]

This leads to a broader discussion about how the electride nature of these 69 candidates was overlooked in the literature. We note that most of the papers reporting on these compounds only discuss synthesis conditions and crystallography. No electron distribution characterization or *ab initio* computing is usually provided and therefore, the electride character of the compound could not be detected. It is interesting though to observe that many of the electrides are reported as Zintl phases (at least 13 of the 69), which are defined as compounds of element groups I-II bonded with elements from groups III-VII.[33] Such species can defy rules of common oxidation states by the coordination of multiple anions together in order to satisfy their own electron counting rules.[34] However, it can be seen in the structures of our candidates that they do *not* exhibit anion aggregation and possess a structural dependence on electron density to reduce repulsion between cations. We believe that many compounds reported as Zintl phases to explain non-matching oxidation states might be, in fact, electrides. An alternative explanation often reported relates to the presence of unidentified $H^-$ in the structure. This has explicitly been claimed in the literature for the $Mn_5Si_3$ class of compounds,[35] and is seen for example with $A_5X_3H$, in which H can stabilize ternary phases that do not exist for the corresponding $A_5X_3$ binaries.[36] It is impossible to discount the presence of H in the original experimental structures, but we offer evidence to support our assignment as electrides. The first is that, according to DFT, all compounds in this study are stable or almost stable versus decomposition to competing phases (see SI). The second is that we successfully recover known electrides. Finally, the presence of H in compounds can be removed by chemical treatment to arrive at the species we compute. Interestingly, the easy uptake and release of H from electrides has been deemed to be essential to the catalytic activity of $Ca_2N$.[37]

We also take advantage of our large dataset to revisit the established design principles for electrides, *i.e.* the presence of a strongly electropositive element, non-matching oxidation states and cavities in the structure. There is a consistent inclusion of electropositive elements in our candidates but broader in scope than has been considered in the past. While the most common group is the alkaline earth metals, the rare-earth elements and alkaline metals are also represented, it is interesting to note that only 1 candidate contains Li. Perhaps this principle can even be extended to the exclusion of electronegative elements as there are zero chalcogenide- or halogen-containing compounds in our candidates. Instead we see nitrides ($N^{3-}$), silicides ($Si^{4-}$), or germanides ($Ge^{4-}$) as common anions.

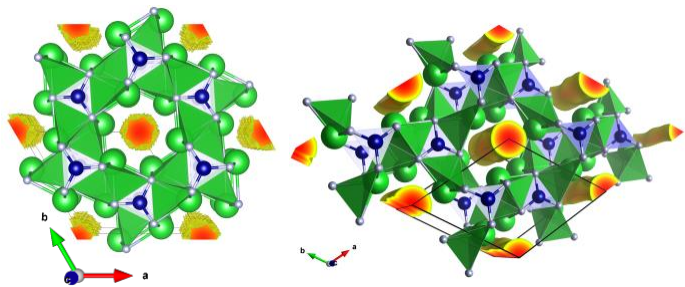

*Figure 4: 1D electride behaviour in $Ba_3CrN_3$ and $Sr_3CrN_3$. Ba/Sr (green), Cr (blue), N (white).*

It can be seen that the most popular group of candidates in this study do not follow the rule of excess electron in the sum of oxidation states. We note however that in many cases oxidation state assignment can be extremely ambiguous.[38] Moreover, there are many materials with apparently non-matching oxidation states but that are not electrides, such as $La_3Te_4$.[39] This can be seen more broadly in the materials project database; there are 751 cases of existing composition in the ICSD which have oxidation states that would sum to +1 based on commonly accepted ion charge states, from $Ag_2F$ to YS (excluding f-block elements). These compounds have 1,141 phases, of which 469 have calculated band gaps greater than 0.05 eV and therefore are extremely unlikely to be electrides (according to Materials Project database). In a nutshell, non-matching oxidation states can be used to explain the electride nature of a compound but is not an efficient predictor of electrides *a priori*.

The necessity for cavities in the crystal system is also corroborated by our results but simple consideration of large voids is insufficient. Looking at the bonded plane perpendicular to the 1D electron density in $Ca_5Sb_3$, as an example, it is possible to see that the anionic electron resides in one of the smaller pores of the structure (see Figure 5) as the electron distribution allows the cations to draw closer together. Indeed, the cation-electron distance is even smaller than the cation-anion distance. Similarly, in the structure for $Sr_5Si_3$ (Figure 3) the bonds between the cation and the anion are 3.30 Å and 3.43 Å whereas the distances between the cation and electron are 2.68 and 2.77 Å. This concept has been arrived upon for high-pressure electrides previously, whereby it was observed that pore space for anionic electron occupation "only becomes important when the size of the enclosed space becomes comparable to the size of typical atoms".[40] These distances indicate that the anionic electrons are volumetrically rather small, which is surprising as there is no positive nucleus contracting the electron distribution about a point in space. Such a behavior is problematic in terms of independent identification of electrides from structure alone, as 3D structure vacancies of the order of the materials bond length or smaller are ubiquitous in materials science.

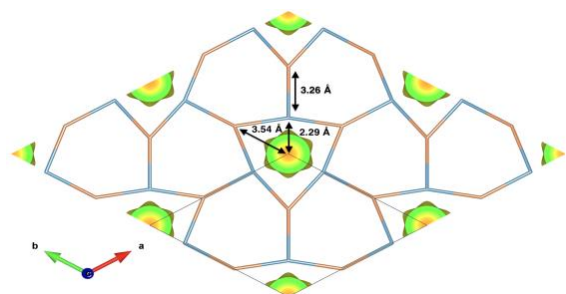

*Figure 5: relevant inter-species distances for bonded plane of $Ca_5Sb_3$ showing the pore size of the electron.*

Finally, while in this work the candidates are grouped based on structure, structure does not necessarily define electride behavior. In our study $K_3Rb$ and $Ca_3Tl$ have identical structures and formula-types but present different electron densities around $E_F$. $K_3Rb$ shows a diffuse distribution whereas $Ca_3Tl$ exhibits very clear 1D electride behavior, as shown in Figure 6. This can also be seen in the different anion-electron coordination environments found for identical structures (see SI).

This paper discusses fewer than half of the total number of candidates. However, we believe that we have adequately illustrated that the true extent of electride materials has until now been incredibly limited and refer the reader to the SI for electronic structure results and valence electron density presented on an individual basis. Further details can be found here, for example, three candidates ($Nd_3In$, $Sr_5Bi_3$ and $Sr_{11}Mg_2Si_{10}$) exhibit band splitting for spin-up and spin-down states, which is evidence of bulk magnetism.

In summary, we find that while the rules of electride identification can be instructive, they ought not limit the consideration of new and existing materials. We propose that many candidates have long been known but difficulties in observing electron density around the Fermi level in experiments mean electrides have been overlooked in the past. As such, high-throughput studies such as this are critical for uncovering powerful new functionalities of solid state systems for a wide variety of applications. By revisiting systems not considered in this way before, we find new electrides that do not have oxidation states indicating an excess electron, that contain redox active transition metals and have narrow cavities in the crystal structure. Our method successfully recovers 2D, 1D and 0D electride systems in their native state, significantly broadening their applicability. Combined with such a large number of potential candidates, we show that this class of industry relevant materials is much wider in scope and accessibility than ever illustrated before. It is expected that the compounds discussed in this paper will be of importance for a wide variety of applications and the fundamental understanding of materials.

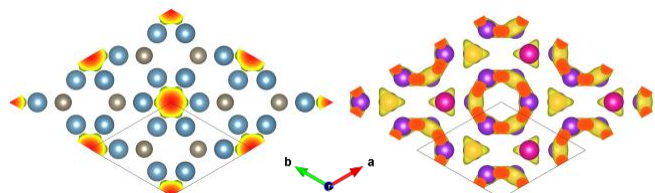

*Figure 6: top-down electron density within the same range of the Fermi energy for compounds with identical structure and formula type $Ca_3Tl$ (left) and $K_3Rb$ (right).*

## AUTHOR INFORMATION

**Corresponding Author**

E-mail: geoffroy.hautier@uclouvain.be
Tel: +32 10 47 35 01

## ACKNOWLEDGMENT

L.A.B. is a beneficiary of a "MOVE-IN Louvain" Incoming Post-doctoral Fellowship, co-funded by the Marie Curie Actions of the European Commission and F.R. was supported by the F.R.S.-FNRS project HTBaSE (contract no PDR-T.1071.15). Computer resources were provided by the Université catholique de Louvain (CISM/UCL) and the Consortium des Equipements de Calcul Intensif en Fédération Wallonie Bruxelles (CECI) funded by the F.R.S.-FNRS. G.-M. R. is grateful to the F.R.S.-FNRS for financial support.

# High-Throughput Identification of Electrides from all Known Inorganic Materials- Supplementary Information

Lee A. Burton, Francesco Ricci, Gian-Marco Rignanese, Geoffroy Hautier.

Herein, the structures and band diagrams for the full list of 69 candidate electrides are displayed in order of the Materials Project materials ID number. Red and blue colour of the bands is used to indicate spin up/down energies, where band splitting can be indicative of magnetism. The chemical formula, crystal type and symmetry are also included, along with origin of the structure information (e.g. ICSD etc). Using the structure_matcher functionality of the pymatgen analysis module the structures are grouped together by nominal compound 'type' nomenclature, for example "CrB-type". Finally, we scale back the isosurface to find an 'origin' of the anionic electron and construct a coordinate environment for each case from this.

| MPID | Formula | Energy above hull (eV/atom) | Providence | System | Symmetry symbol | Symmetry number | Electride type |
|---|---|---|---|---|---|---|---|
| mp-10045 | $Ba_5As_3$ | 0.0085 | ICSD | hexagonal | $P6_3/mcm$ | 193 | 1D |
| mp-10961 | $La_5Si_3$ | 0 | ICSD | tetragonal | $I4/mcm$ | 140 | 0D |
| mp-11404 | PrGa | 0 | ICSD | orthorhombic | Cmcm | 63 | 2D |
| mp-11840 | BaAg | 0 | ICSD | orthorhombic | Pnma | 62 | 0D |
| mp-12467 | $Ca_5Sb_3$ | 0.00080 | ICSD | hexagonal | $P6_3/mcm$ | 193 | 1D |
| mp-12612 | $Ba_3Ag_2$ | 0.00054 | ICSD | trigonal | R-3 | 148 | 0D |
| mp-12723 | CaAu | 0 | ICSD | orthorhombic | Cmcm | 63 | 2D |
| mp-12905 | $Ba_3CrN_3$ | 0 | ICSD | hexagonal | $P6_3/m$ | 176 | 1D |
| mp-12906 | $Sr_3CrN_3$ | 0 | ICSD | hexagonal | $P6_3/m$ | 176 | 1D |
| mp-13053 | SrSi | 0.0074 | ICSD | orthorhombic | Pnma | 62 | 0D |
| mp-1334 | $Y_2C$ | 0 | ICSD | trigonal | R-3m | 166 | 2D |
| mp-1363 | $Na_2Au$ | 0 | ICSD | tetragonal | $I4/mcm$ | 140 | 1D |
| mp-1448 | NdGa | 0 | ICSD | orthorhombic | Cmcm | 63 | 2D |
| mp-1464 | $Nd_5Ge_3$ | 0 | ICSD | hexagonal | $P6_3/mcm$ | 193 | 1D |
| mp-1563 | CaSi | 0 | ICSD | orthorhombic | Cmcm | 63 | 2D |
| mp-15698 | $Sr_5As_3$ | 0 | ICSD | hexagonal | $P6_3/mcm$ | 193 | 1D |
| mp-1698 | SrSn | 0 | ICSD | orthorhombic | Cmcm | 63 | 2D |
| mp-1730 | BaGe | 0 | ICSD | orthorhombic | Cmcm | 63 | 2D |
| mp-17325 | $Ba_5Sn_3$ | 0.0092 | ICSD | tetragonal | $I4/mcm$ | 140 | 0D |
| mp-17720 | $Sr_5Sn_3$ | 0.0030 | ICSD | tetragonal | $I4/mcm$ | 140 | 0D |
| mp-17757 | $Sr_5Ge_3$ | 0.0025 | ICSD | tetragonal | $I4/mcm$ | 140 | 0D |
| mp-18167 | $Ca_3Cd_2$ | 0 | ICSD | tetragonal | $P4_2/mnm$ | 136 | 0D |
| mp-18316 | $Mg_2Pd$ | 0 | ICSD | cubic | Fd-3m | 227 | 0D |
| mp-1884 | $Ca_5Ge_3$ | 0 | ICSD | tetragonal | $I4/mcm$ | 140 | 0D |
| mp-20909 | $La_3In$ | 0 | ICSD | cubic | Pm-3m | 221 | 0D |
| mp-2147 | SrGe | 0 | ICSD | orthorhombic | Cmcm | 63 | 2D |
| mp-21483 | $Nd_3In$ | 0 | ICSD | cubic | Pm-3m | 221 | 0D |
| mp-2360 | CaGe | 0 | ICSD | orthorhombic | Cmcm | 63 | 2D |
| mp-24153 | $LaH_2$ | 0 | ICSD | cubic | Fm-3m | 225 | 0D |
| mp-2499 | BaSi | 0 | ICSD | orthorhombic | Cmcm | 63 | 2D |
| mp-2538 | $Y_5Si_3$ | 0 | ICSD | hexagonal | $P6_3/mcm$ | 193 | 1D |
| mp-2585 | $Sr_5Sb_3$ | 0 | ICSD | hexagonal | $P6_3/mcm$ | 193 | 1D |
| mp-2631 | $Ba_4Al_5$ | 0 | ICSD | hexagonal | $P6_3/mmc$ | 194 | 0D |
| mp-2661 | SrSi | 0 | ICSD | orthorhombic | Cmcm | 63 | 2D |
| mp-2686 | $Ca_2N$ | 0 | ICSD | trigonal | R-3m | 166 | 2D |
| mp-28489 | $Ca_5(GaN_2)_2$ | 0 | ICSD | orthorhombic | Cmca | 64 | 0D |
| mp-29620 | $Sr_5Bi_3$ | 0 | ICSD | hexagonal | $P6_3/mcm$ | 193 | 1D |
| mp-29621 | $Ba_5Bi_3$ | 0 | ICSD | hexagonal | $P6_3/mcm$ | 193 | 1D |
| mp-30355 | SrAg | 0 | ICSD | orthorhombic | Pnma | 62 | 0D |
| mp-30357 | $Sr_3Ag_2$ | 0 | ICSD | trigonal | R-3 | 148 | 0D |
| mp-30366 | $Ca_3Au$ | 0 | ICSD | orthorhombic | Pnma | 62 | 0D |
| mp-30367 | $Ca_5Au_2$ | 0 | ICSD | monoclinic | C2/c | 15 | 0D |
| mp-30368 | $Ca_5Au_3$ | 0 | ICSD | tetragonal | $I4/mcm$ | 140 | 0D |
| mp-30422 | $Sr_7Au_3$ | 0 | ICSD | hexagonal | $P6_3mc$ | 186 | 1D |
| mp-371 | $La_3Tl$ | 0 | ICSD | cubic | Pm-3m | 221 | 0D |
| mp-4579 | LaSiRu | 0 | ICSD | tetragonal | P4/nmm | 129 | 2D |

| | | | | | | | | |
|---|---|---|---|---|---|---|---|---|
| **mp-4738** | PrScGe | 0 | ICSD | tetragonal | I4/mmm | 139 | 2D |
| **mp-4854** | NdScGe | 0 | ICSD | tetragonal | I4/mmm | 139 | 2D |
| **mp-542680** | $Na_3In_2Au$ | 0 | ICSD | cubic | Fd-3m | 227 | 0D |
| **mp-542681** | $Na_3In_2Ag$ | 0 | ICSD | cubic | Fd-3m | 227 | 0D |
| **mp-567342** | $Nd_4MgIr$ | 0 | ICSD | cubic | F-43m | 216 | 0D |
| **mp-569535** | $Ca_2Bi$ | 0 | ICSD | tetragonal | I4/mmm | 139 | 0D |
| **mp-570400** | $Ba_7Al_{10}$ | 0 | ICSD | trigonal | R-3m | 166 | 0D |
| **mp-573908** | $Sr_{11}(MgSi_5)_2$ | 0.0061 | ICSD | monoclinic | C2/m | 12 | 0D |
| **mp-605873** | $Pr_4MgRu$ | 0 | ICSD | cubic | F-43m | 216 | 0D |
| **mp-630923** | $Ba_3Pb_5$ | 0 | ICSD | orthorhombic | Cmcm | 63 | 0D |
| **mp-645130** | $Pr_4MgCo$ | 0 | ICSD | cubic | F-43m | 216 | 0D |
| **mp-693** | $Pr_3Tl$ | 0.0038 | ICSD | cubic | Pm-3m | 221 | 0D |
| **mp-7376** | $Sr_3(AlSn)_2$ | 0 | ICSD | orthorhombic | Immm | 71 | 0D |
| **mp-746** | $Sr_5Si_3$ | 0 | ICSD | tetragonal | I4/mcm | 140 | 0D |
| **mp-7507** | $Sr_3Li_2$ | 0 | ICSD | tetragonal | P4_2/mnm | 136 | 0D |
| **mp-793** | $Ca_5Si_3$ | 0 | ICSD | tetragonal | I4/mcm | 140 | 0D |
| **mp-8320** | SmScSi | 0 | ICSD | tetragonal | I4/mmm | 139 | 2D |
| **mp-872** | BaSn | 0 | ICSD | orthorhombic | Cmcm | 63 | 2D |
| **mp-973508** | $KRb_3$ | 0.0063 | OQMD | tetragonal | I4/mmm | 139 | 0D |
| **mp-974066** | $Nd_3Sm$ | 0.0089 | OQMD | cubic | Pm-3m | 221 | 0D |
| **mp-976115** | $K_3Rb$ | 0.0052 | OQMD | hexagonal | P6_3/mmc | 194 | 0D |
| **mp-984744** | $Ca_3Tl$ | 0.003 | OQMD | hexagonal | P6_3/mmc | 194 | 1D |
| **mp-9909** | $Ba_5Sb_3$ | 0 | ICSD | hexagonal | P6_3/mcm | 193 | 1D |

mp-10045; Ba$_5$As$_3$. Symmetry: hexagonal; P6_3/mcm (number 193). Structure: Mn$_5$Si$_3$-type. Origin: ICSD;[2] comments: referred to as intermetallic. Anionic electron coordination environment: octahedral with Ba.

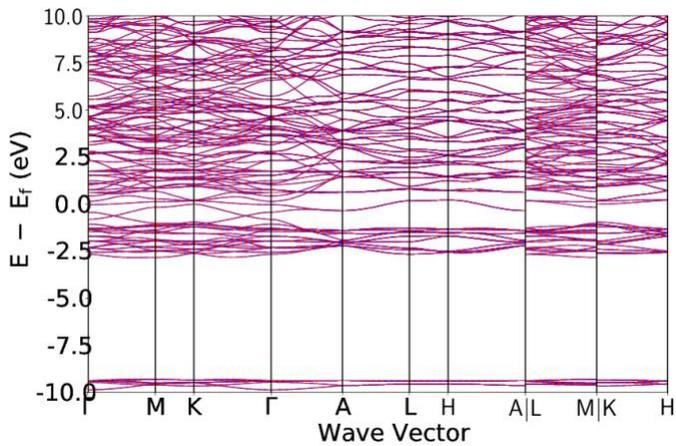
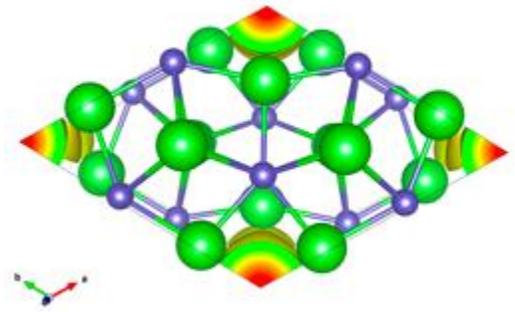

mp-10961; La$_5$Si$_3$. Symmetry: tetragonal; I4/mcm (number 140). Structure: Cr$_5$B$_3$-type. Origin: ICSD;[4] comments: an "electronic delocalization" is observed that is proposed to be the origin of "unusual electrical resistivity" for this compound. Anionic electron coordination environment: tetrahedral with La.

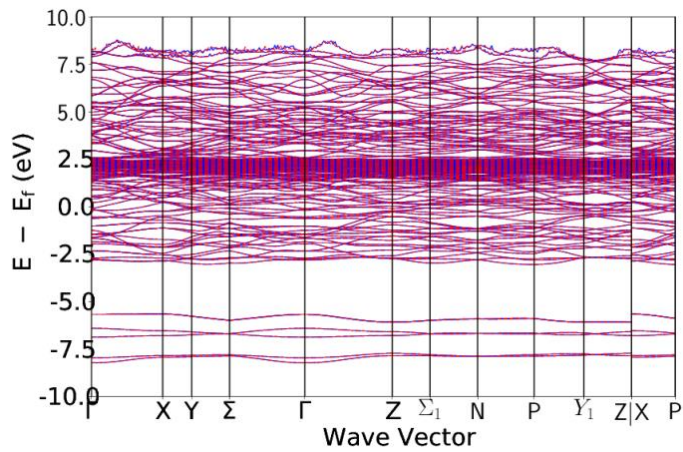
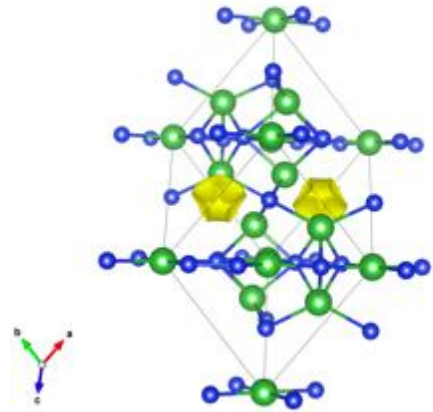

mp-11404; PrGa. Symmetry: orthorhombic; Cmcm (number 63). Structure: CrB-type. Origin: ICSD;[5] comments: referred to as intermetallic. Anionic electron coordination environment: tetrahedral with Pr.

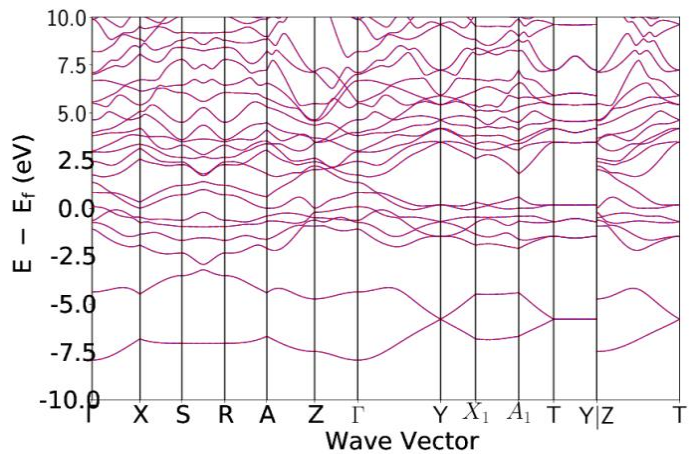
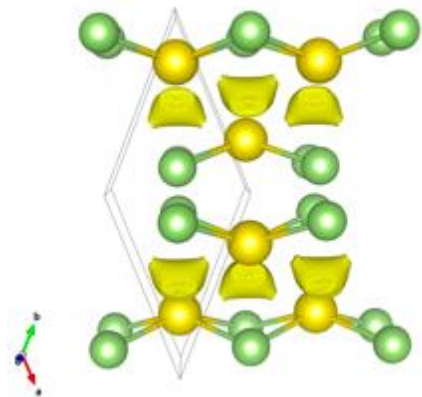

mp-11840; BaAg. Symmetry: Orthorhombic; Pnma (number 62). Structure: FeB-type. Origin: ICSD;[7] comments: "The CrB and FeB types have to be considered the simplest members of a structural family based on the trigonal-prismatic coordination". Anionic electron coordination environment: tetrahedral with Ba.

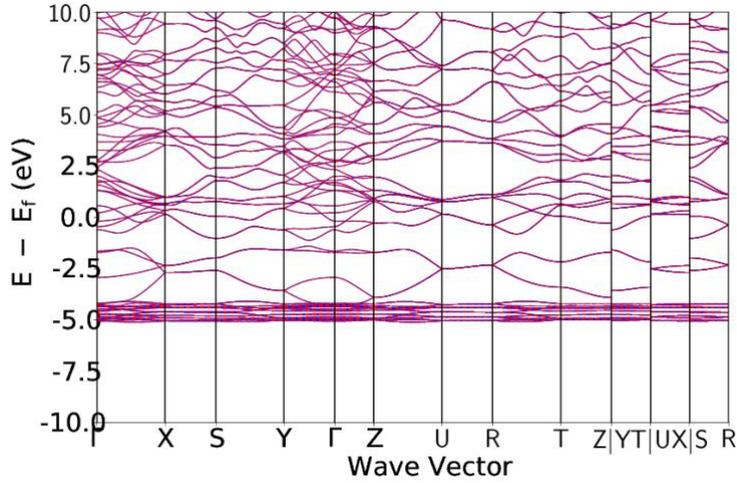 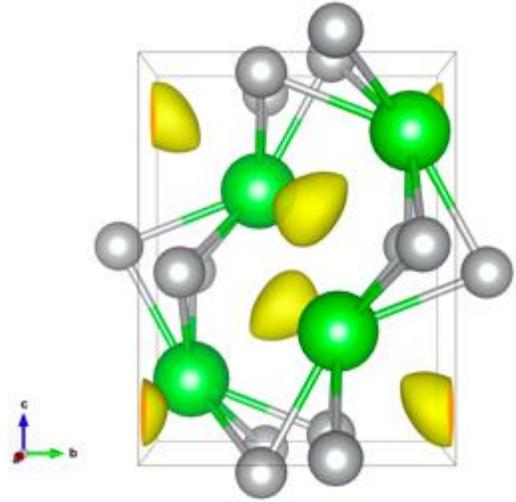

mp-12467; $Ca_5Sb_3$. Symmetry: hexagonal; $P6_3/mcm$ (number 193). Structure: $Mn_5Si_3$-type. Origin: ICSD;[9] comments: referred to as intermetallic. Suggest previous reports of similar structures contain unresolved hydrogen within the 1-d channel of electron density. Anionic electron coordination environment: trigonal planar with Ca.

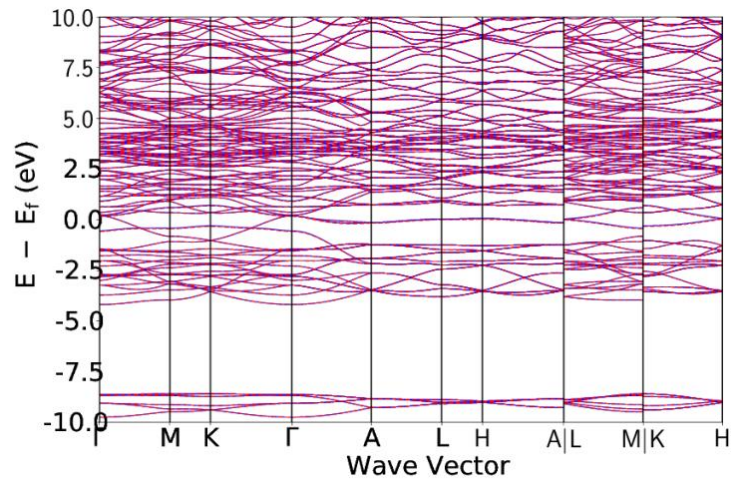 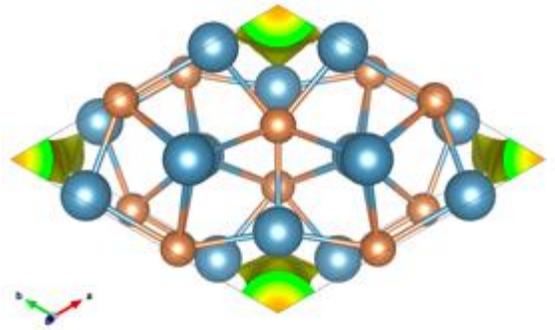

mp-12612; $Ba_3Ag_2$. Symmetry: trigonal; R-3 (number 148). Structure: $Er_3Ni_2$-type. Origin: ICSD;[10] comments: N/A. Anionic electron coordination environment: Octahedral with Ba.

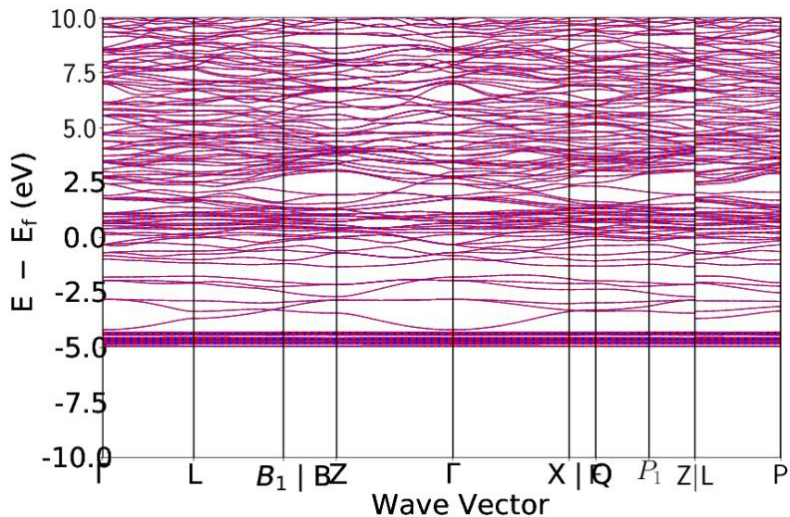 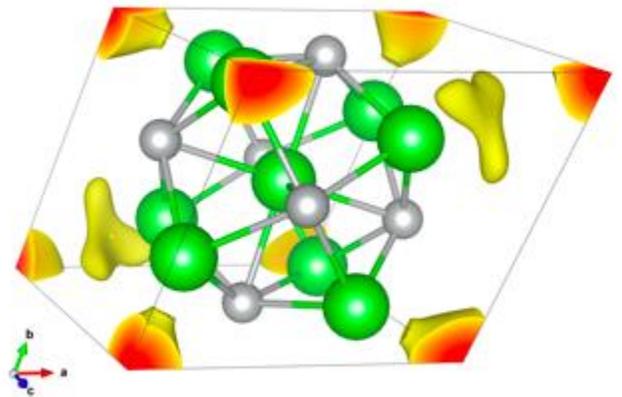

mp-12723; CaAu. Symmetry: orthorhombic; Cmcm (number 63). Structure: CrB-type. Origin: ICSD;[11] comments: N/A. Anionic electron coordination environment: distorted octahedral with Ca and Au.

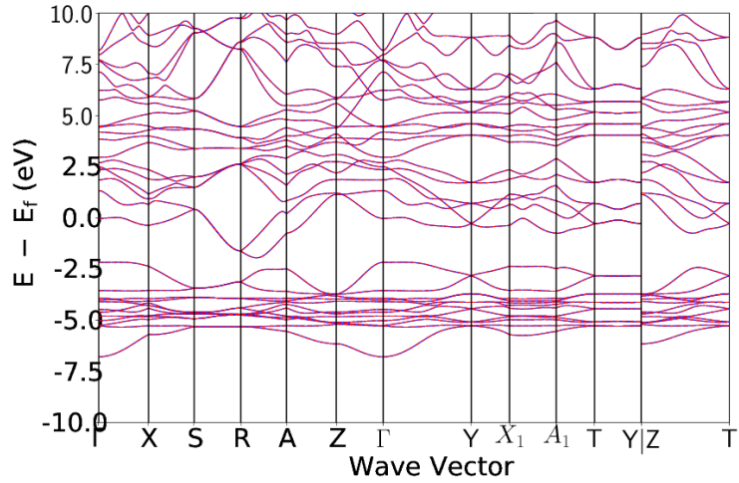
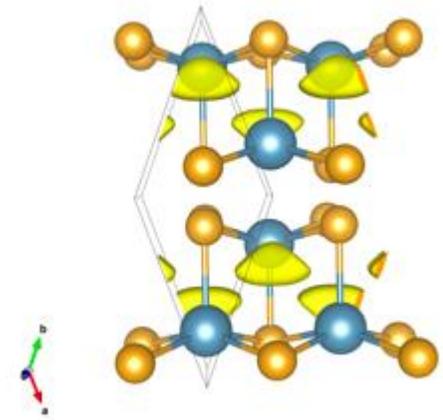

mp-12905; Ba$_3$CrN$_3$. Symmetry: hexagonal; P6$_3$/m (number 176). Structure: Ba$_3$CrN$_3$-type. Origin: ICSD;[12] comments: "The [calculated bond valences for Cr] deviate significantly from the expected oxidation states… and lie invariably closer to 4 than 3". Anionic electron coordination environment: octahedral with Ba.

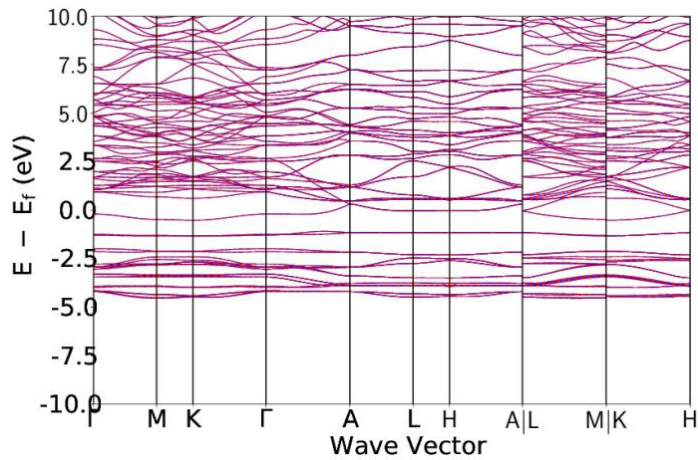
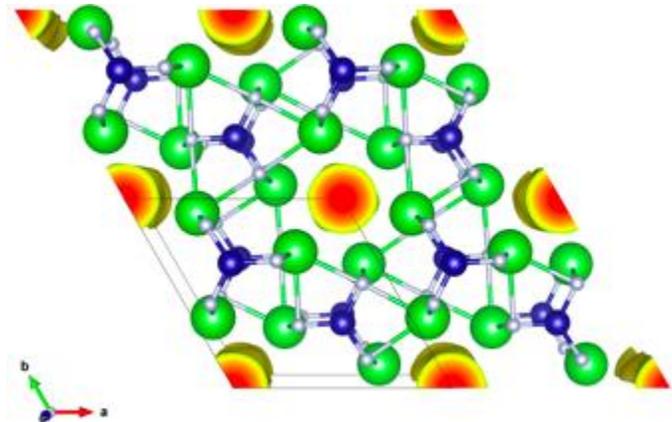

mp-12906; Sr$_3$CrN$_3$. Symmetry: hexagonal; P6$_3$/m (number 176). Structure: Ba$_3$CrN$_3$-type. Origin: ICSD;[12] comments: "The [calculated bond valences for Cr] deviate significantly from the expected oxidation states… and lie invariably closer to 4 than 3". Anionic electron coordination environment: octahedral with Sr.

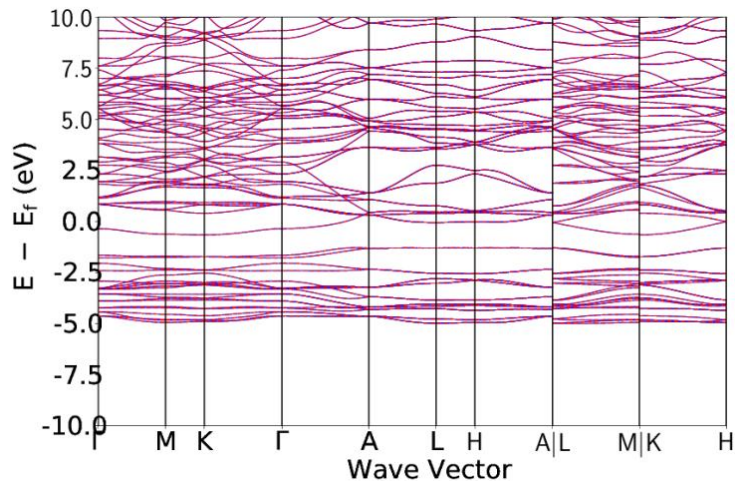
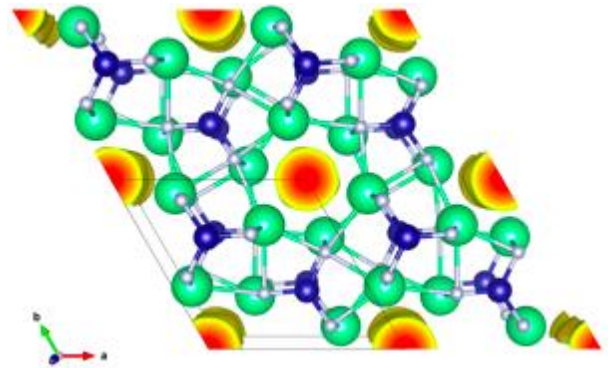

mp-13053; SrSi. Symmetry: Orthorhombic; Pnma (number 62). Structure: FeB-type. Origin: ICSD;[13] *ab inito* predicted structure. Anionic electron coordination environment: distorted octahedral with Sr and Si.

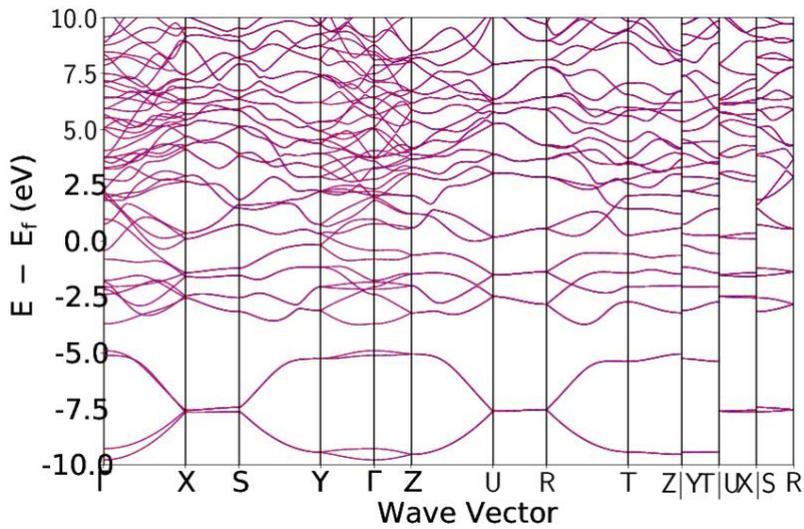
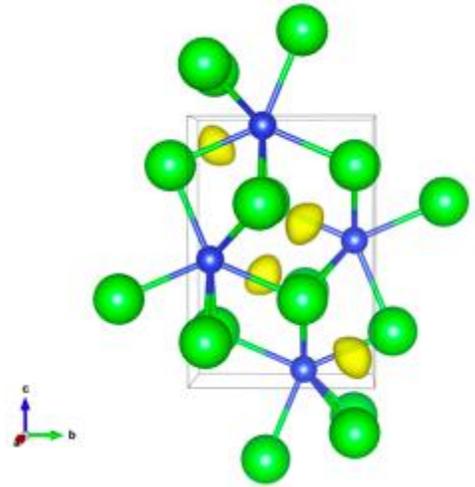

mp-1334; Y$_2$C. Symmetry: trigonal; R-3m (number 166). Structure: anti-CdCl$_2$-type. Origin: ICSD;[15] comments: discussed Y$_2$C as use for hydrogen absorption. Anionic electron coordination environment: trigonal pyramidal with Y.

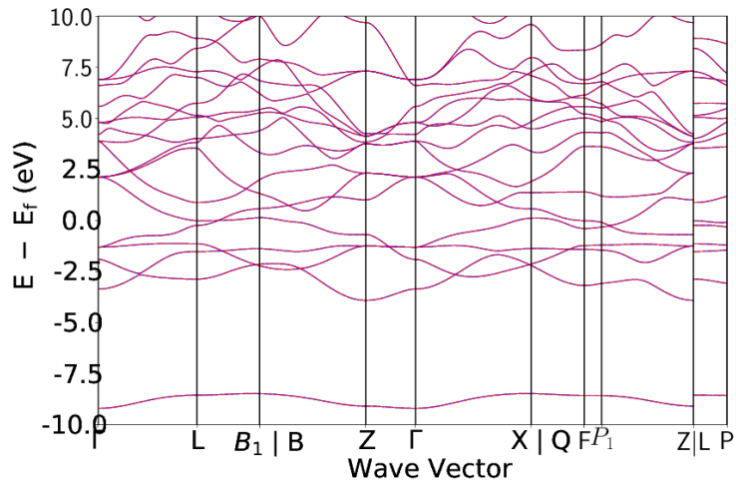
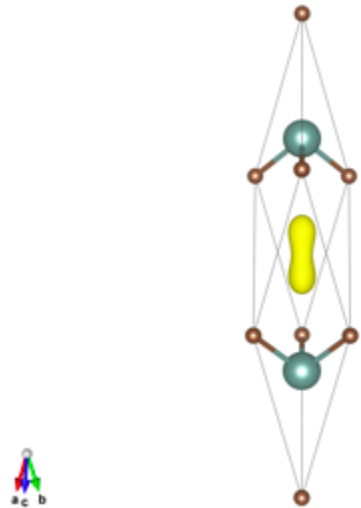

mp-1363; Na$_2$Au. Symmetry: tetragonal; I4/mcm (number 140). Structure: Cu$_2$Mg-type. Origin: ICSD;[16] comment: N/A. Anionic electron coordination environment: tetrahedral with Na.

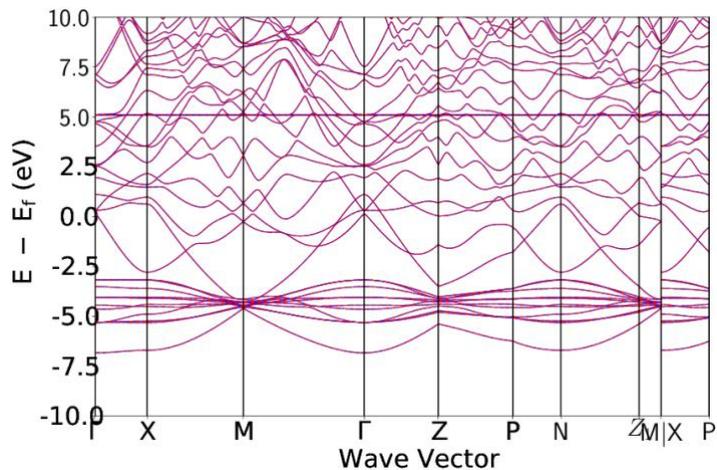
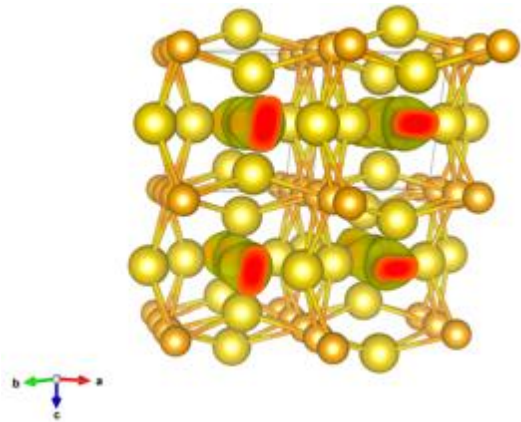

mp-1448; NdGa. Symmetry: orthorhombic; Cmcm (number 63). Structure: CrB-type. Origin: ICSD;[18] comments: referred to as alloy. Anionic electron coordination environment: tetrahedral with Nd.

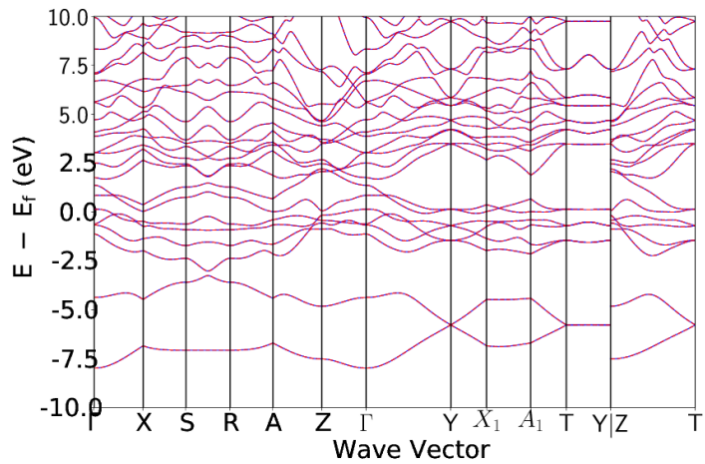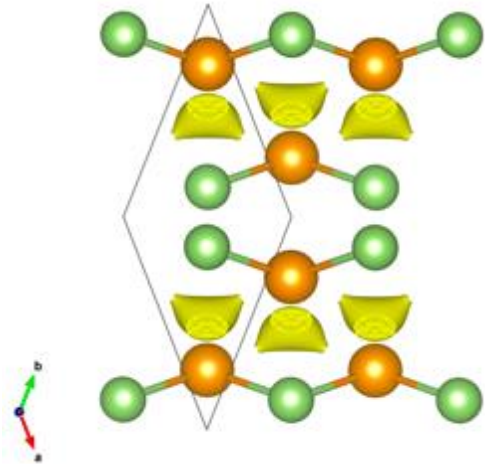

mp-1464; Nd$_5$Ge$_3$. Symmetry: hexagonal; P6_3/mcm (number 193). Structure: Mn$_5$Si$_3$-type. Origin: ICSD;[19] comments: N/A. Anionic electron coordination environment: trigonal pyramidal with Nd.

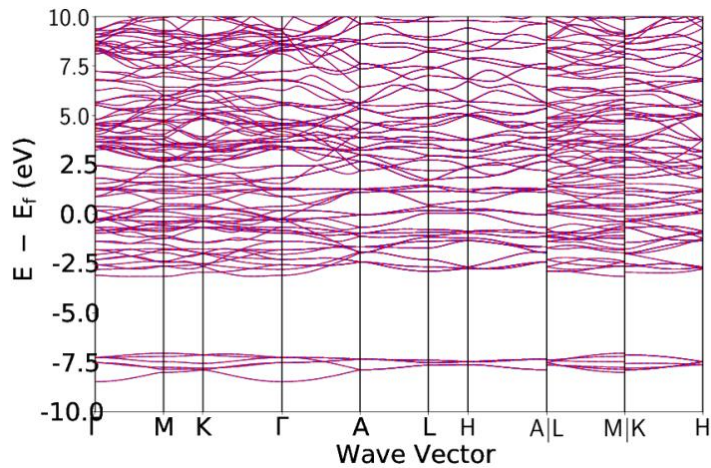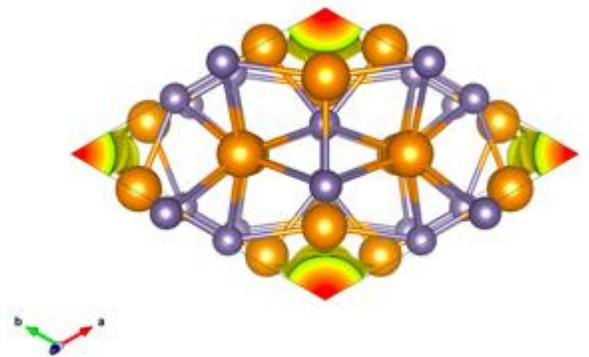

mp-1563; CaSi. Symmetry: orthorhombic; Cmcm (number 63). Structure: CrB-type. Origin: ICSD;[20] comments: high pressure phase, *ab inito* predicted structure. CaSi described as 'Zintl compound'. Electron localization function shows anion covalency 'proving Zintl concept'. ELF also shows 'lone pair regions' in structure. High pressure can induce bonding interactions between cations in similar structures. Anionic electron coordination environment: distorted octahedral with Ca and Si.

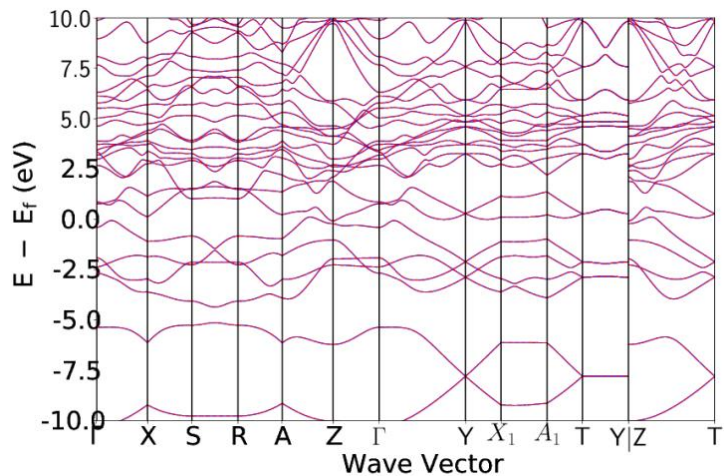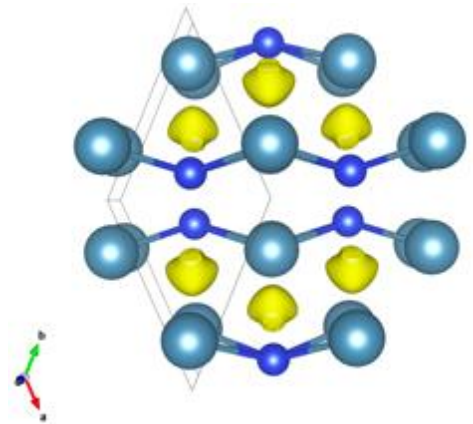

mp-15698; Sr$_5$As$_3$. Symmetry: hexagonal; P6_3/mcm (number 193). Structure: Mn$_5$Si$_3$-type. Origin: ICSD; [2] comments: referred to as intermetallic. Anionic electron coordination environment: trigonal pyramidal with Sr.

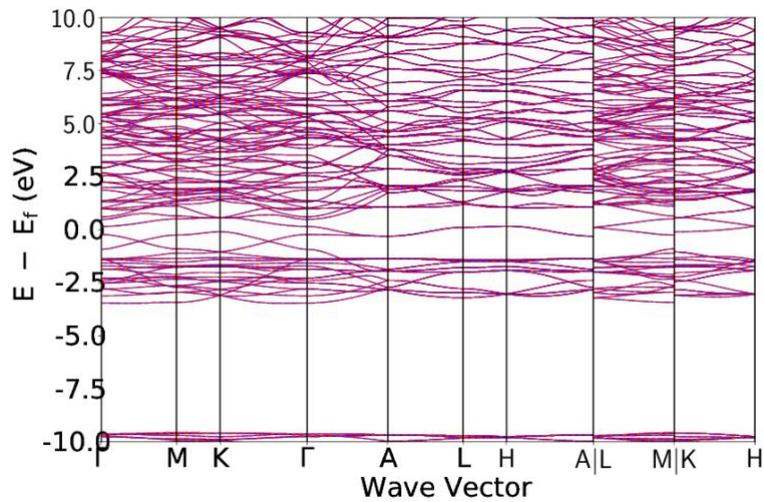
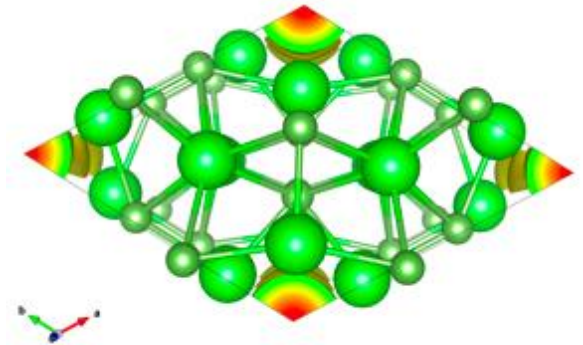

mp-1698; SrSn. Symmetry: orthorhombic; Cmcm (number 63). Structure: CrB-type. Origin: ICSD;[21] comments: N/A. Anionic electron coordination environment: distorted octahedral with Sr and Sn.

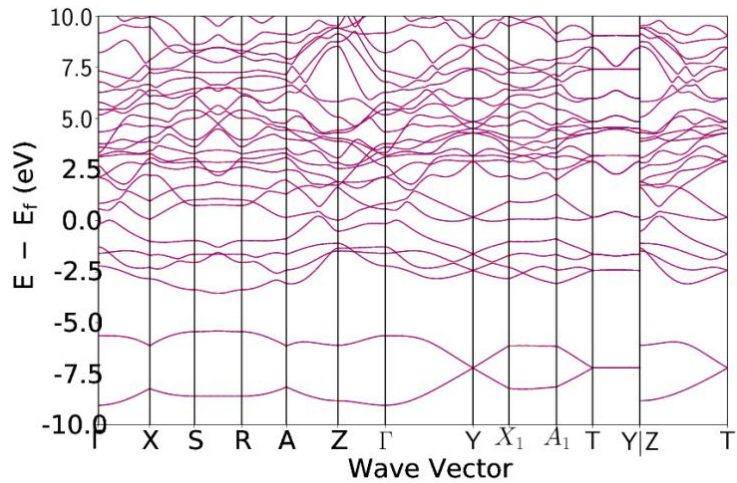
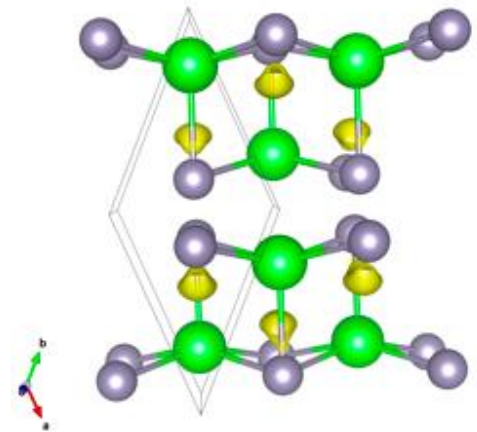

mp-1730: BaGe. Symmetry: orthorhombic; Cmcm (number 63). Structure: CrB-type. Origin: ICSD;[22] comments: described as Zintl: "general valence equation suggests the presence of anion-anion chains", reacts violently with moisture. Anionic electron coordination environment: distorted octahedral with Ba and Ge.

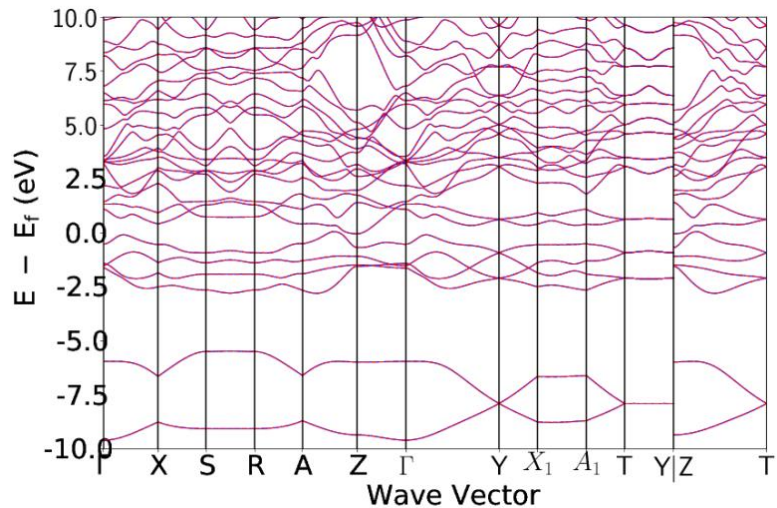
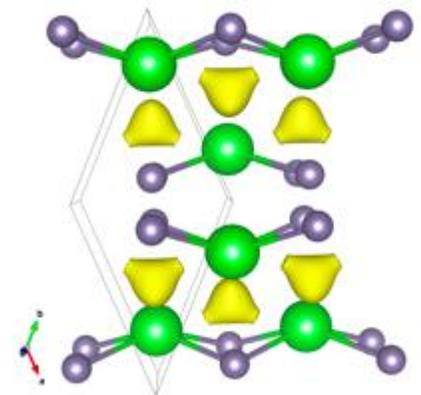

mp-17325; Ba$_5$Sn$_3$. Symmetry: tetragonal; I4/mcm (number 140). Structure: Cr$_5$B$_3$-type. Origin: ICSD;[23] comments: referred to as intermetallic. Anionic electron coordination environment: tetrahedral with Ba.

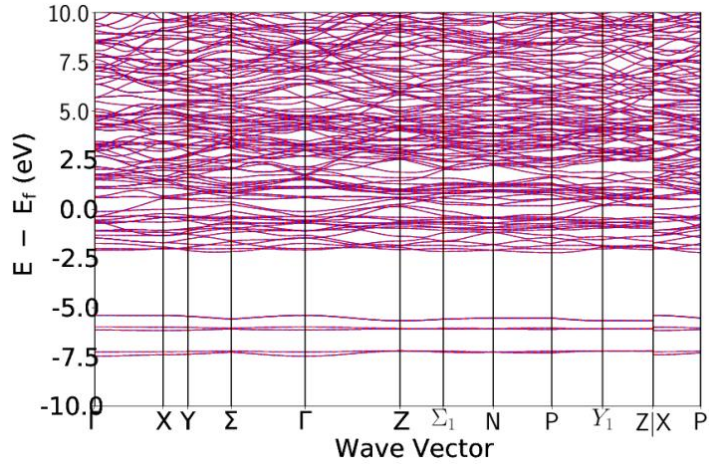
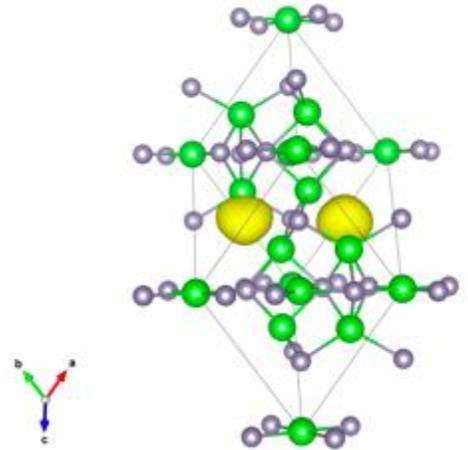

mp-17720; Sr$_5$Sn$_3$. Symmetry: tetragonal; I4/mcm (number 140). Structure: Cr$_5$B$_3$-type. Origin: ICSD;[23] comments: referred to as intermetallic. Anionic electron coordination environment: tetrahedral with Sr.

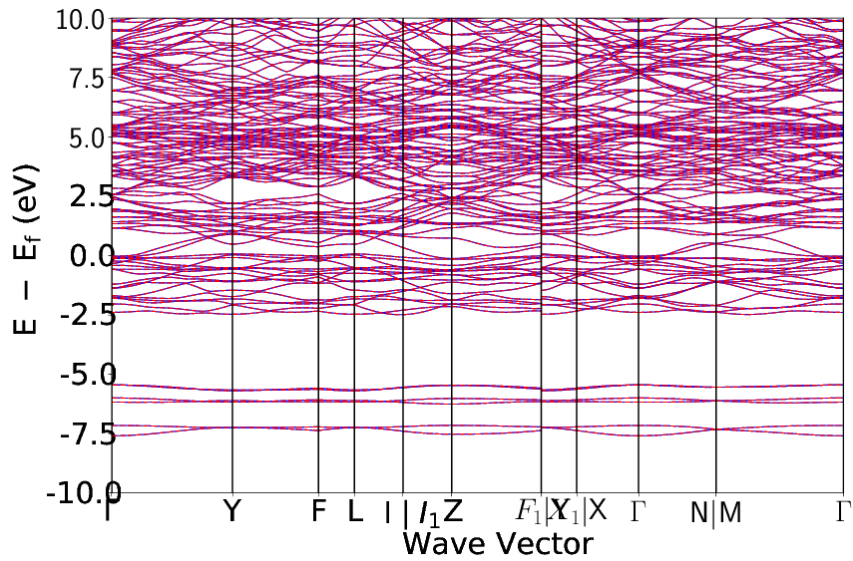
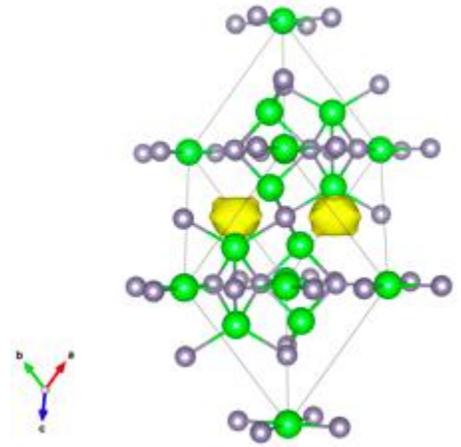

mp-17757; Sr$_5$Ge$_3$. Symmetry: tetragonal; I4/mcm (number 140). Structure: Cr$_5$B$_3$-type. Origin: ICSD;[24] comments: referred to as Zintl compound. Anionic electron coordination environment: tetrahedral with Sr.

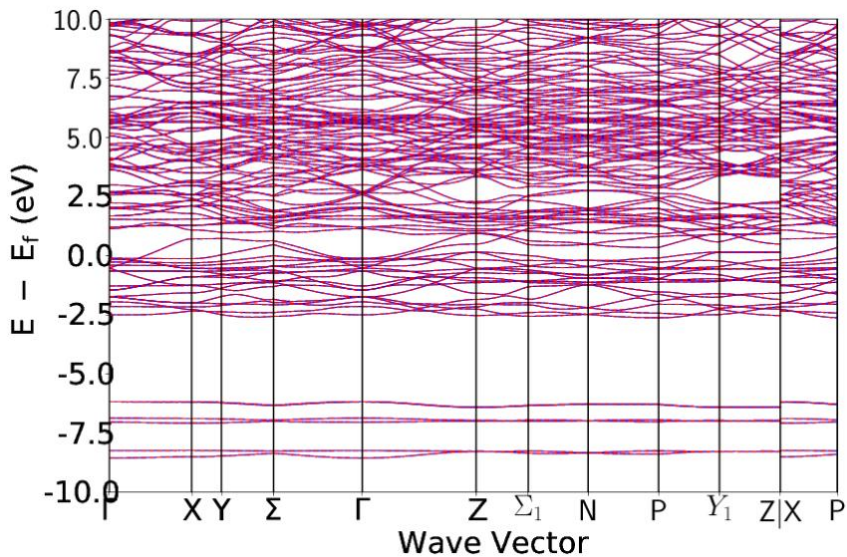
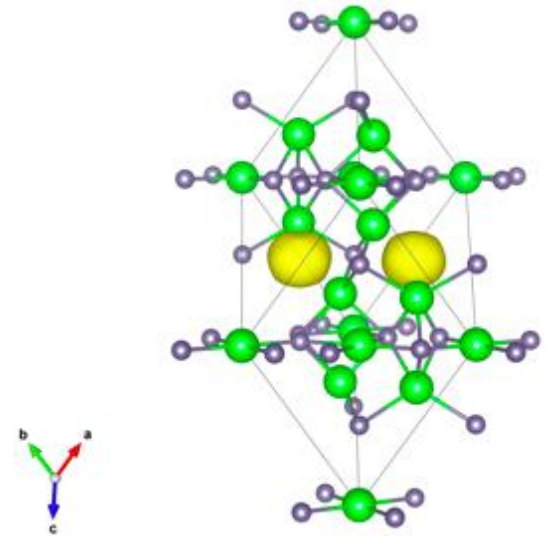

mp-18167; Ca$_3$Cd$_2$. Symmetry: tetragonal; P4_2/mnm (number 136). Structure: Gd$_3$Al$_2$-type. Origin: ICSD;[25] comments: referred to as intermetallic. Anionic electron coordination environment: tetrahedral with Ca.

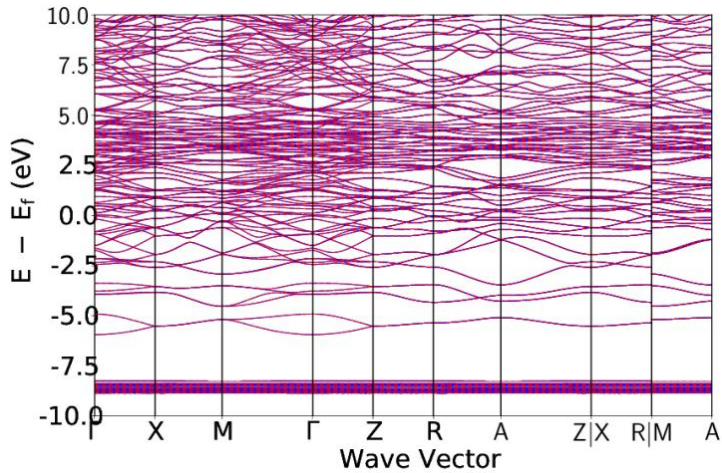 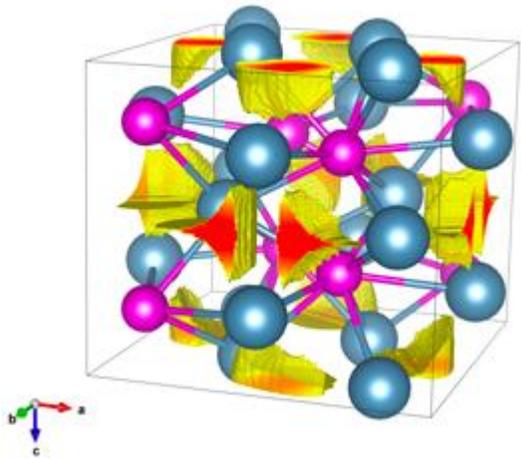

mp-18316; Mg$_2$Pd. Symmetry: cubic; Fd-3m (number 227). Structure: NiTi$_2$-type. Origin: ICSD;[26] comments: N/A. Anionic electron coordination environment: octahedral with Mg.

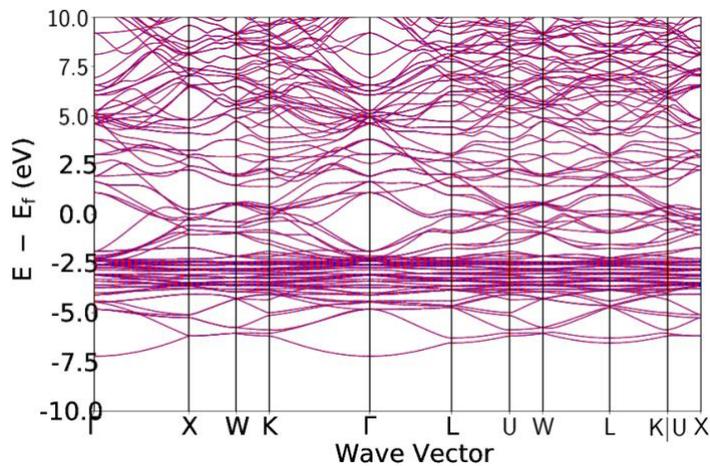 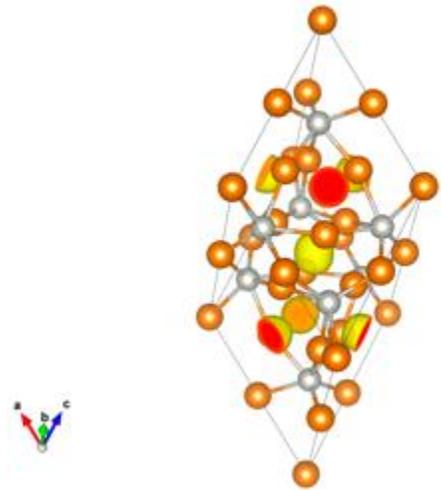

mp-1884; Ca$_5$Ge$_3$. Symmetry: tetragonal; I4/mcm (number 140). Structure: Cr$_5$B$_3$-type. Origin: ICSD;[27] comments: referred go as Zintl phase. Suggest unresolved hydrogen might be included in structure. Anionic electron coordination environment: tetrahedral with Ca.

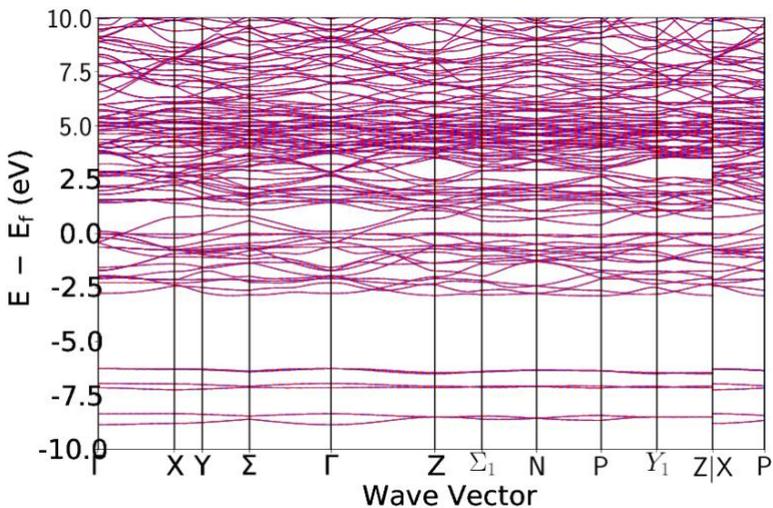 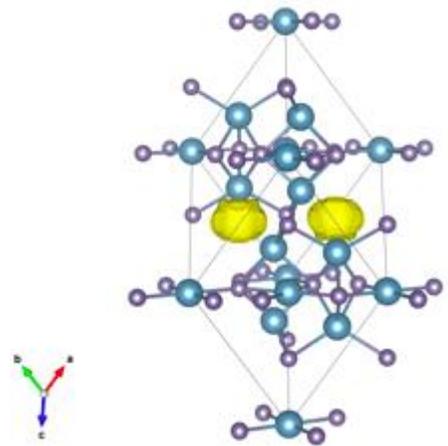

mp-20909; La$_3$In. Symmetry: cubic; Pm-3m (number 221). Structure: Cu$_3$Au-type. Origin: ICSD;[28] comments: superconductor at low temperature. Anionic electron coordination environment: octahedral with La and In.

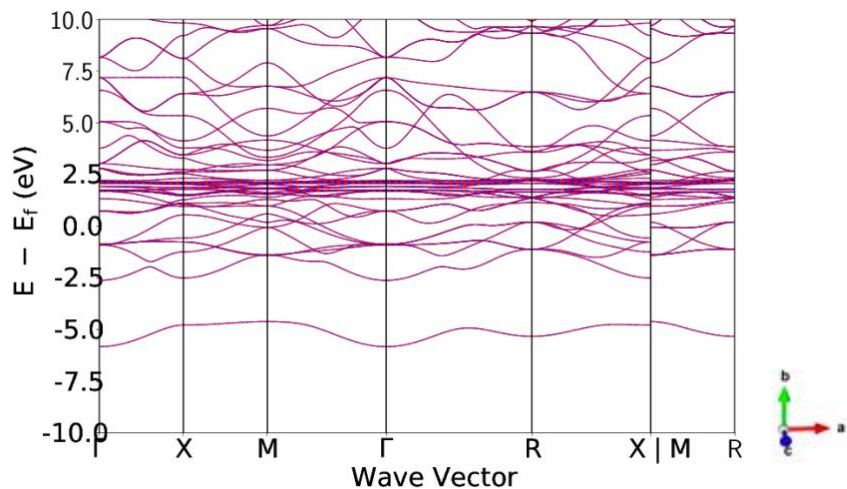
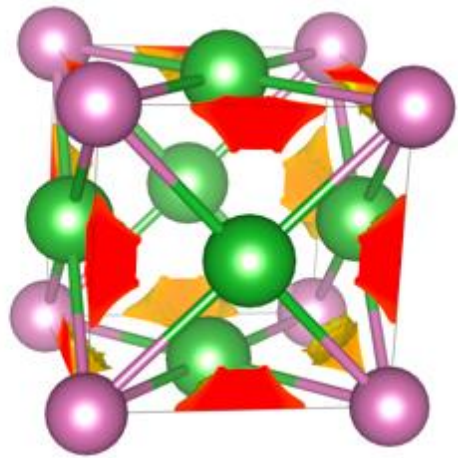

mp-2147; SrGe. Symmetry: orthorhombic; Cmcm (number 63). Structure: CrB-type. Origin: ICSD;[22] comments: referred to as intermetallic. Anionic electron coordination environment: distorted octahedral with Sr and Ge.

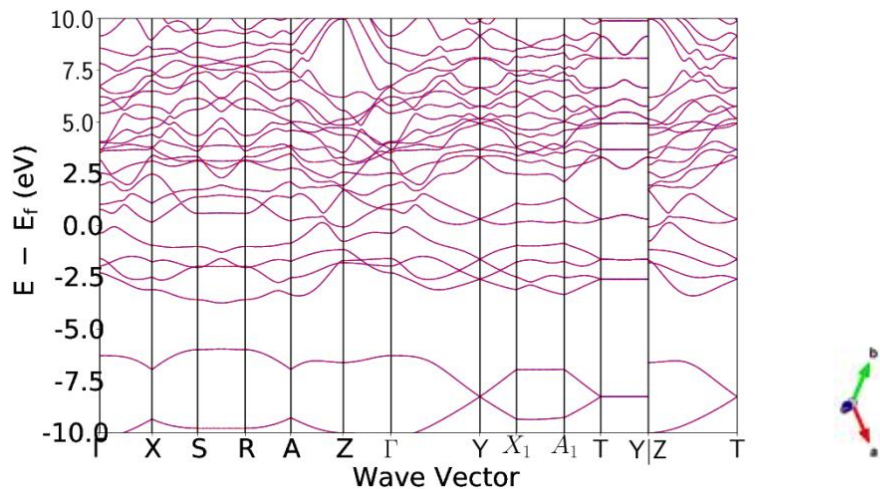
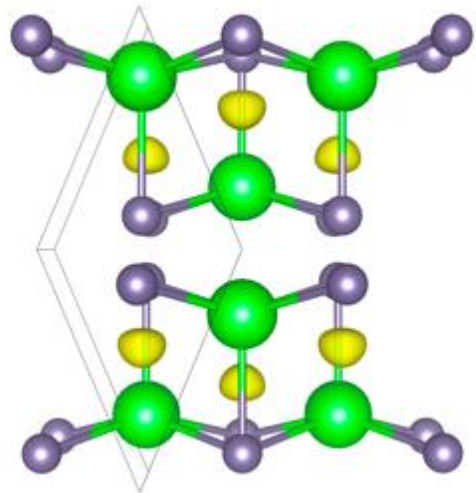

mp-21483; Nd$_3$In. Symmetry: cubic; Pm-3m (number 221). Structure: Cu$_3$Au-type. Origin: ICSD;[29] comments: band splitting indicative of magnetism. Anionic electron coordination environment: octahedral with Nd and In.

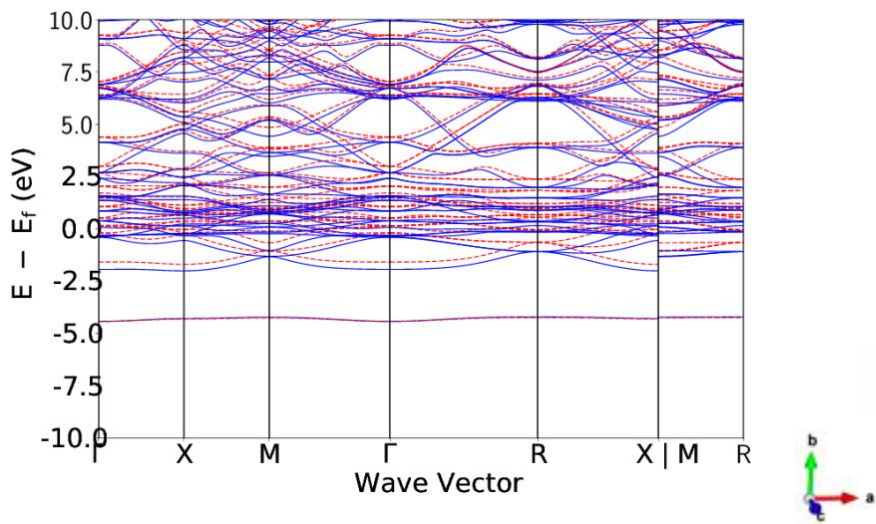
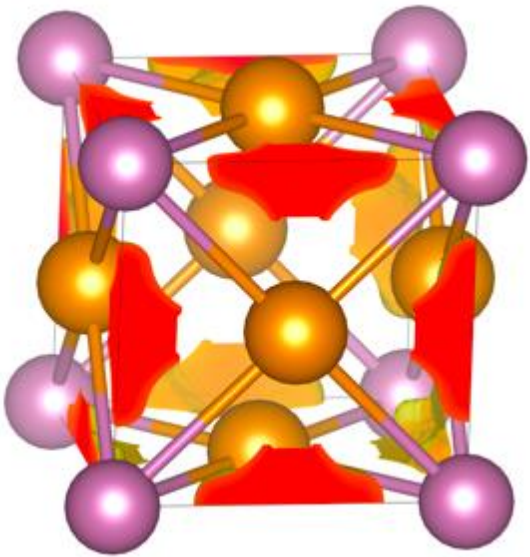

mp-2360; CaGe. Symmetry: orthorhombic; Cmcm (number 63). Structure: CrB-type. Origin: ICSD;[30] comments: N/A. Anionic electron coordination environment: distorted octahedral with Ca and Ge.

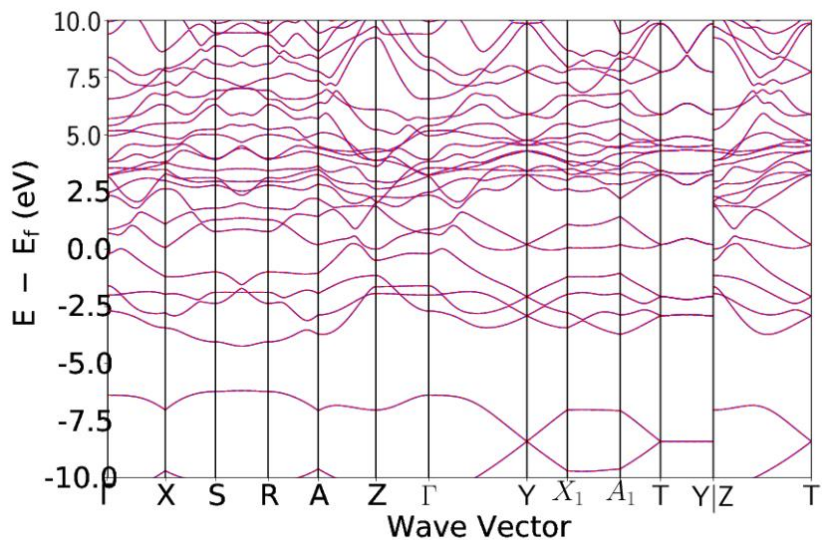 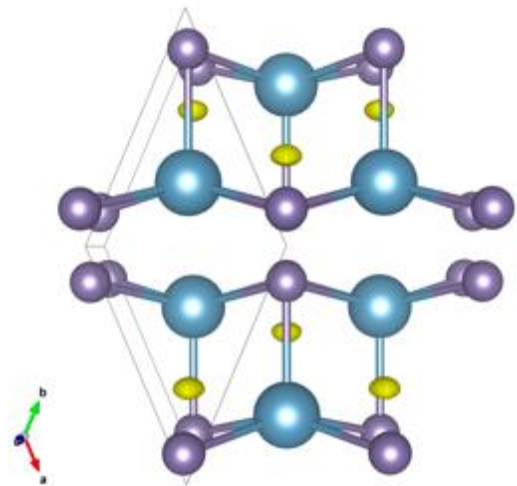

mp-24153; LaH$_2$. Symmetry: cubic; Fm-3m (number 225). Structure: CaF$_2$-type. Origin: ICSD;[31] comments: N/A. Anionic electron coordination environment: square-antiprismatic with H.

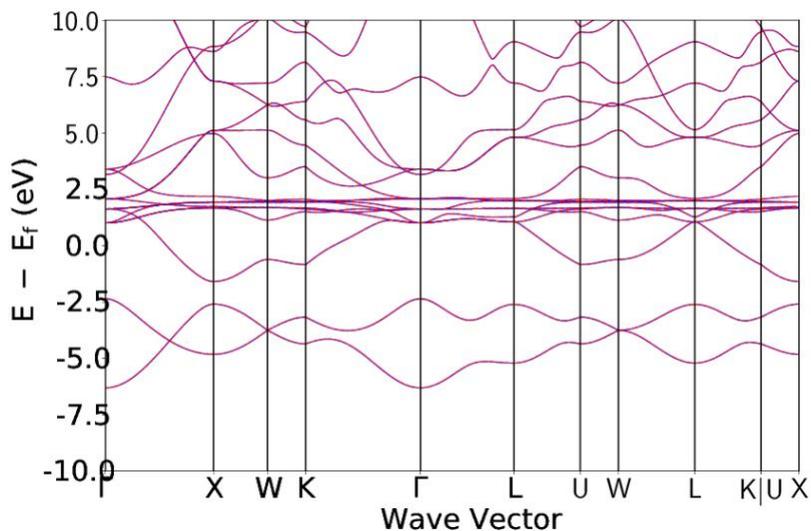 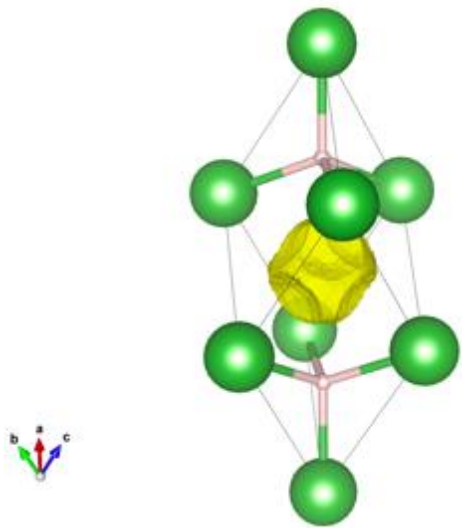

mp-2499; BaSi. Symmetry: orthorhombic; Cmcm (number 63). Structure: CrB-type. Origin: ICSD;[22] comments: referred to as intermetallic. Anionic electron coordination environment: tetrahedral with Ba.

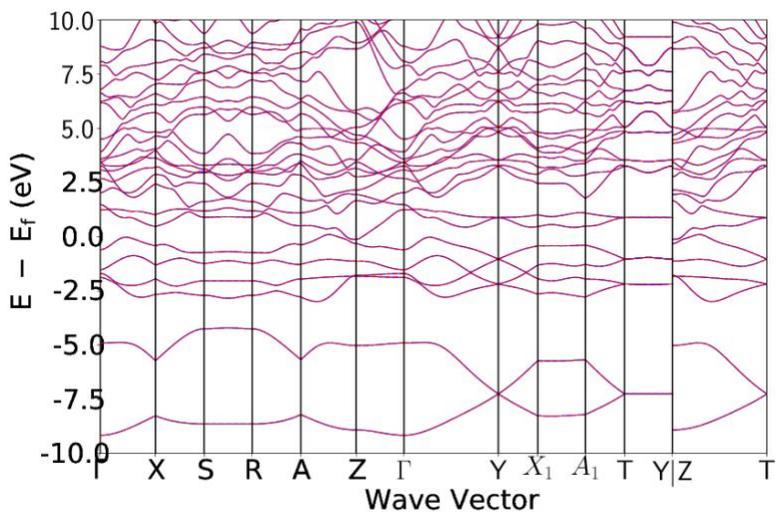 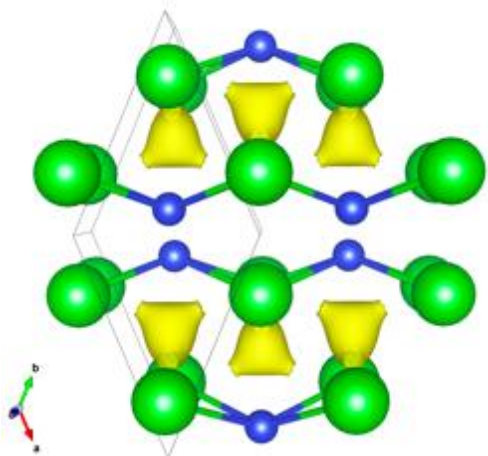

mp-2538; Y₅Si₃. Symmetry: hexagonal; P6_3/mcm (number 193). Structure: Mn₅Si₃-type. Origin: ICSD;[32] comments: Mn5Si3-type compounds can be "stuffed" with variety of elements without change in lattice type. "Stuffed" compounds referred to as "Nowotny" phases. Y₅Si₃ successfully subject to boron intercalation. Anionic electron coordination environment: octahedral with Y.

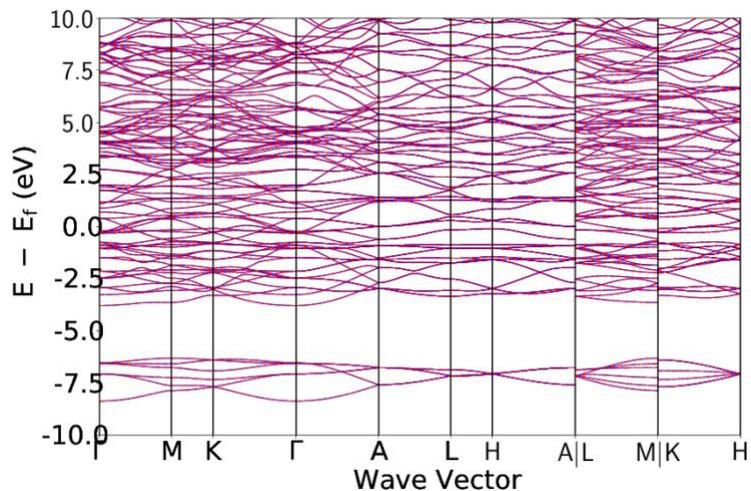 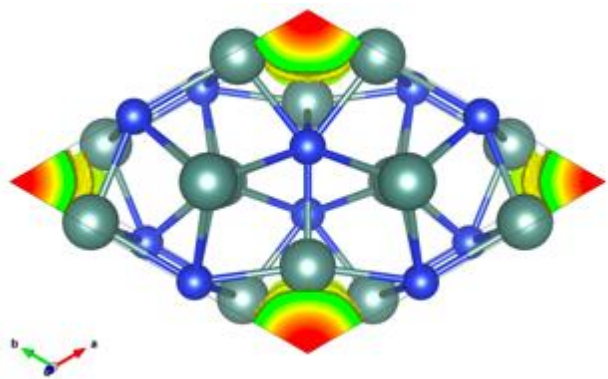

mp-2585; Sr₅Sb₃. Symmetry: hexagonal; P6_3/mcm (number 193). Structure: Mn₅Si₃-type. Origin: ICSD;[33] comments: referred to as Zintl. Anionic electron coordination environment: trigonal planar with Sr.

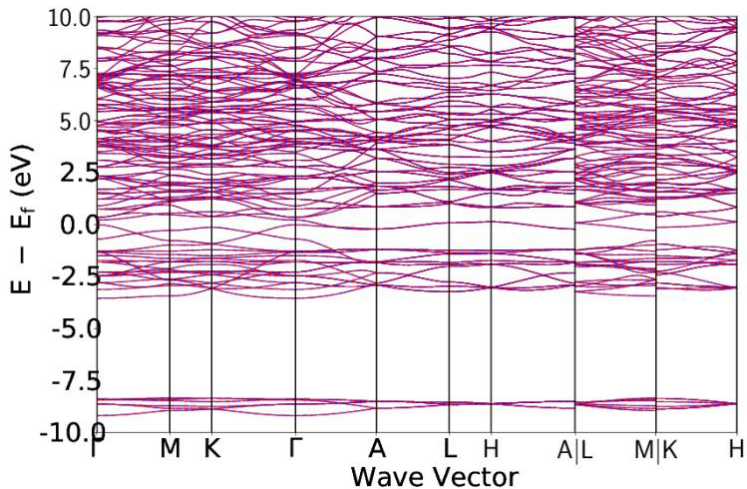 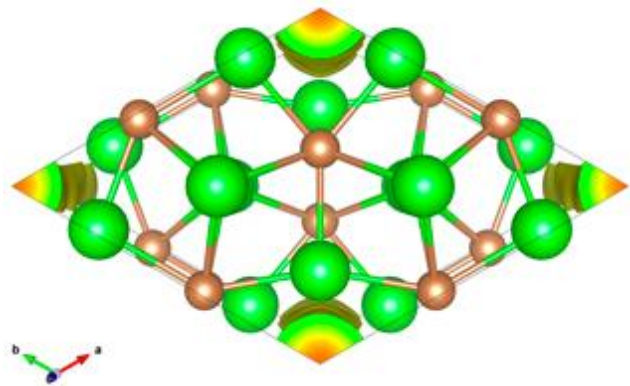

mp-2631; Ba₄Al₅. Symmetry: hexagonal; P6_3/mmc (number 194). Structure: Ba₄Al₅-type. Origin: ICSD;[34] comments: referred to as Zintl compound. "Al atoms the polyanion should bear a charge of minus 7 which means that the number of counter cations is… too large in Ba₄Al₅". "Possible hydrogen impurities can be ruled out". Anionic electron coordination environment: linear with Ba.

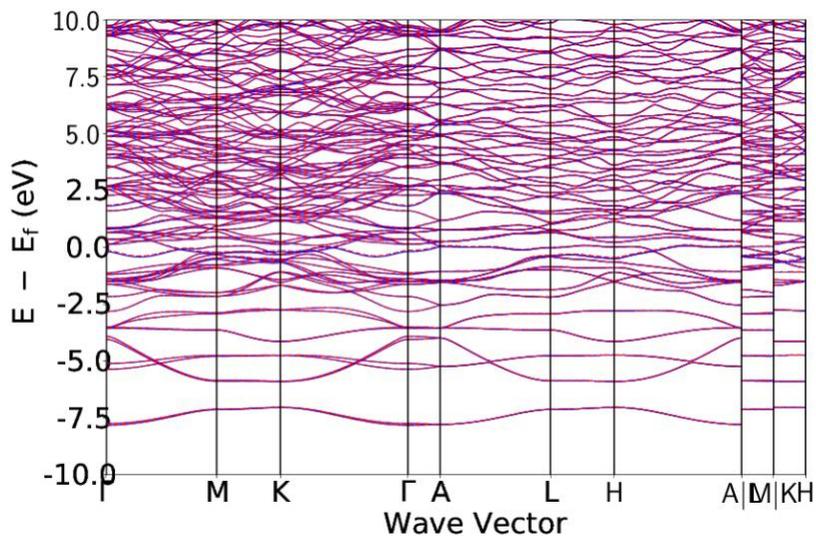 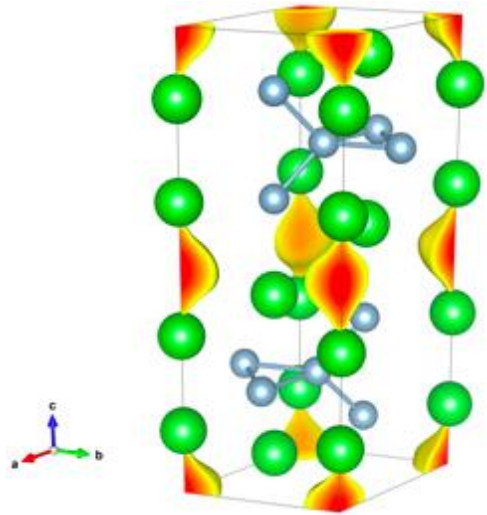

mp-2661; SrSi. Symmetry: orthorhombic; Cmcm (number 63). Structure: CrB-type. Origin: ICSD;[13] *ab inito* predicted structure. Anionic electron coordination environment: distorted octahedral with Sr and Si.

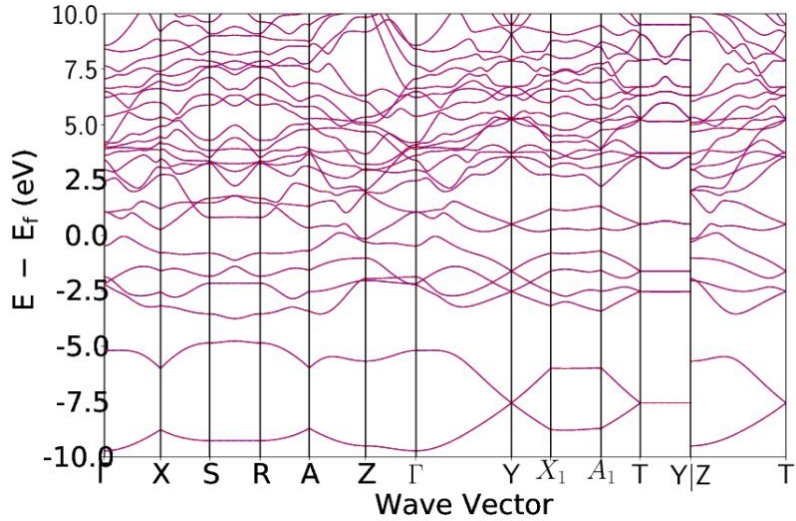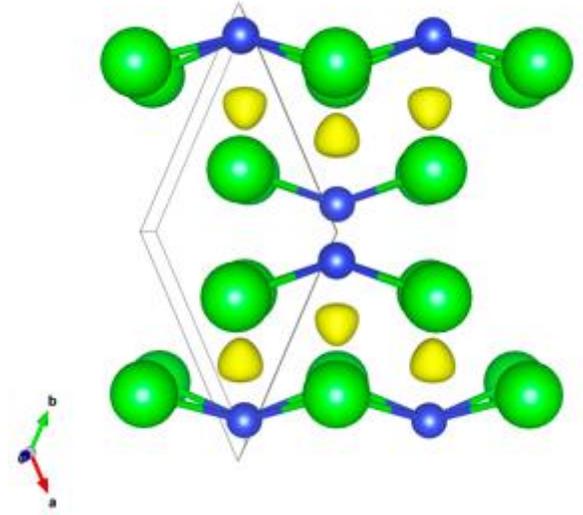

mp-2686; Ca$_2$N. Symmetry: trigonal; R-3m (number 166). Structure: anti-CdCl$_2$-type. Origin: ICSD;[35] comments: describes 'metallic inter-layer bonding' of 'nearly free remaining electrons'. Refers to as 'void metal'. 'the metallically bonded regions are more compressible than the ionically bonded ones… compressibility along the *c*-axis was found to be about twice as high compared to that along the *a*-axis'. Anionic electron coordination environment: octahedral with Ca.

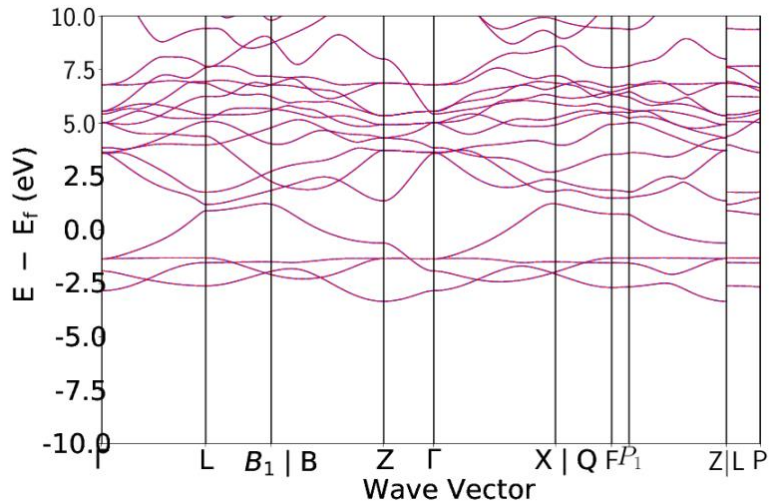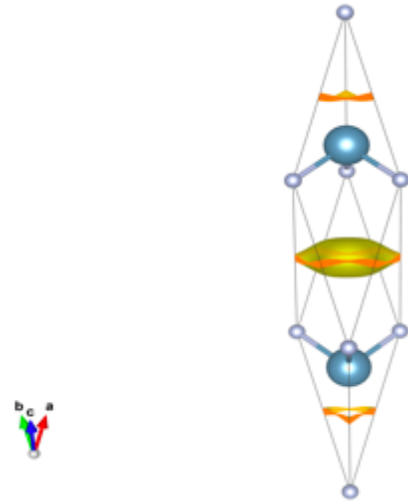

mp-28489; Ca$_5$(GaN$_2$)$_2$. Symmetry: orthorhombic; Cmca (number 64). Structure: Ca$_5$(GaN$_2$)$_2$-type. Origin: ICSD;[36] comments: referred to as Zintle phase. Anionic electron coordination environment: octahedral with Ca.

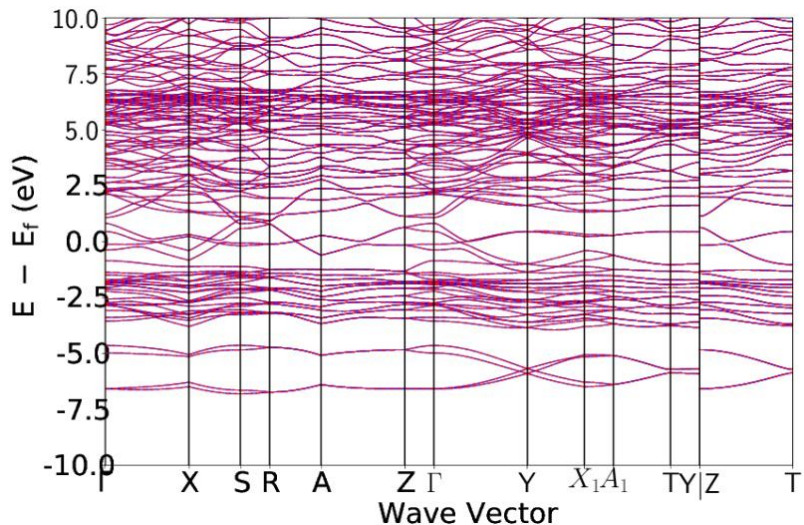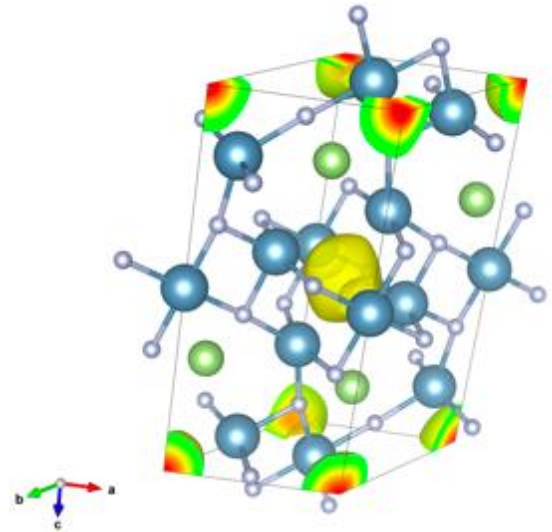

mp-29620; Sr₅Bi₃. Symmetry: hexagonal; P6_3/mcm (number 193). Structure: Mn₅Si₃-type. Origin: ICSD;[37] comments: band splitting indicative of magnetism. Anionic electron coordination environment: trigonal planar with Sr.

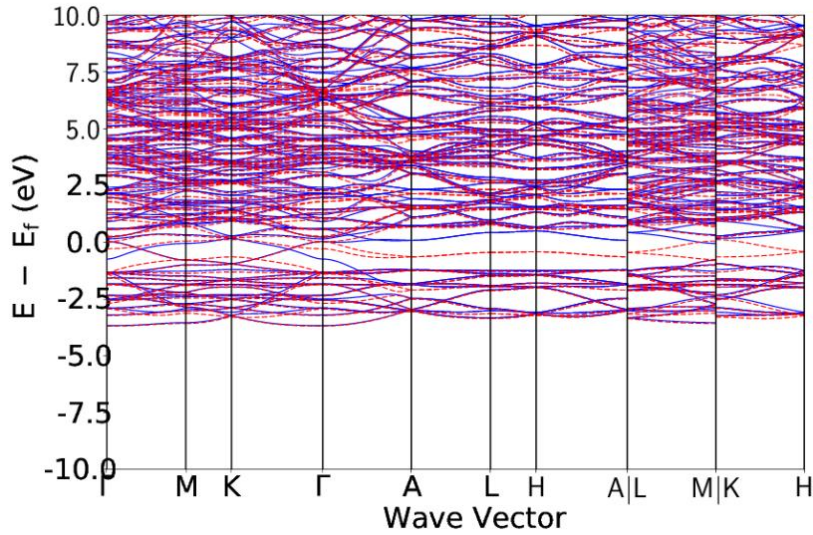
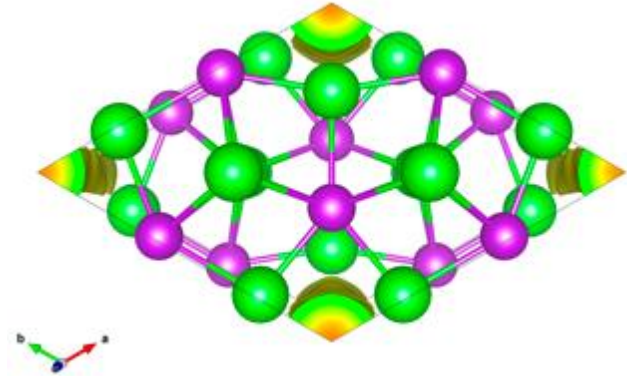

mp-29621; Ba₅Bi₃. Symmetry: hexagonal; P6_3/mcm (number 193). Structure: Mn₅Si₃-type. Origin: ICSD;[37] comments: N/A. Anionic electron coordination environment: octahedral with Ba.

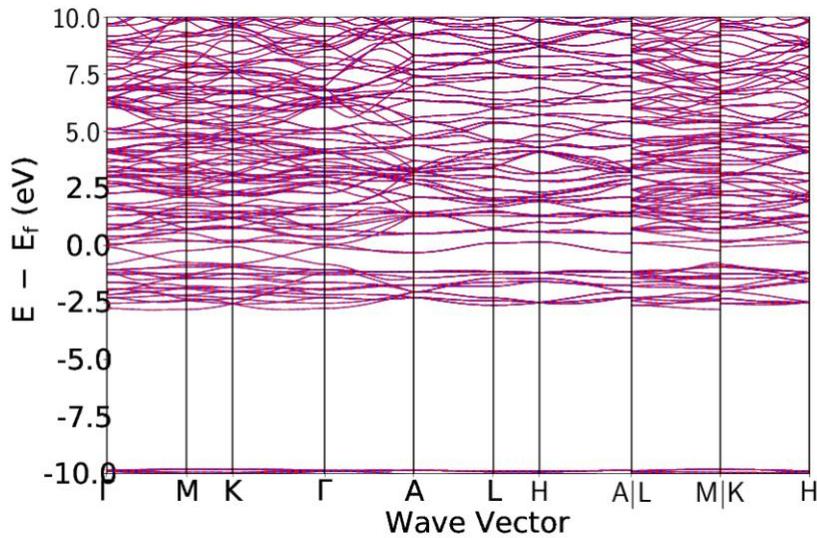
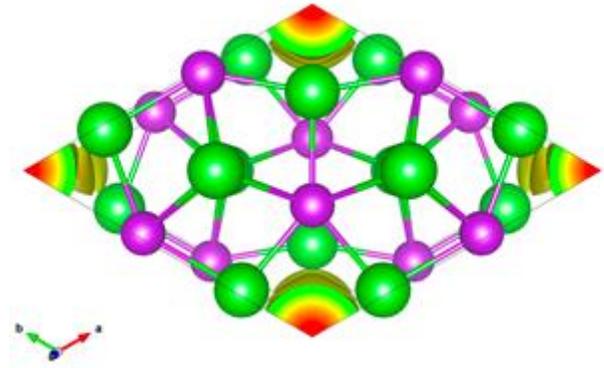

mp-30355; SrAg. Symmetry: Orthorhombic; Pnma (number 62). Structure: SrAg-type. Origin: ICSD;[7] comments: N/A. Anionic electron coordination environment: distorted octahedral with Sr and Ag.

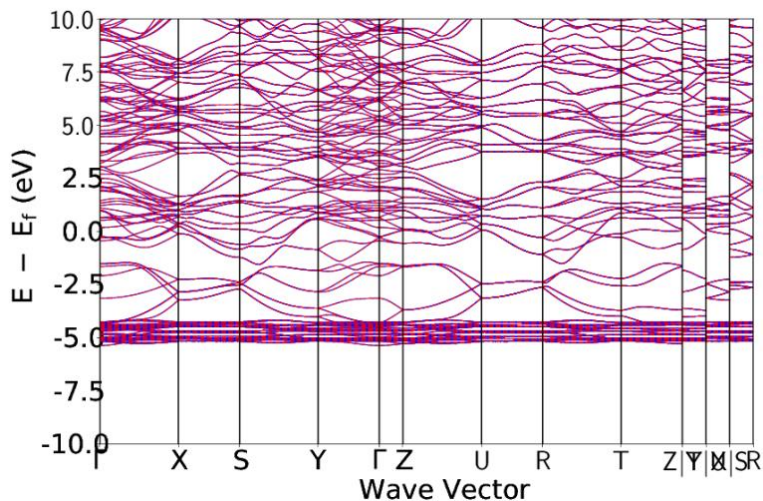
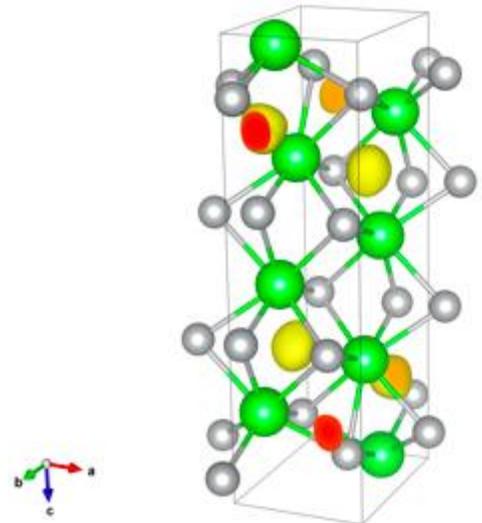

mp-30357; $Sr_3Ag_2$. Symmetry: trigonal; R-3 (number 148). Structure: $Er_3Ni_2$-type. Origin: ICSD;[38] comments: N/A. Anionic electron coordination environment: octahedral with Sr.

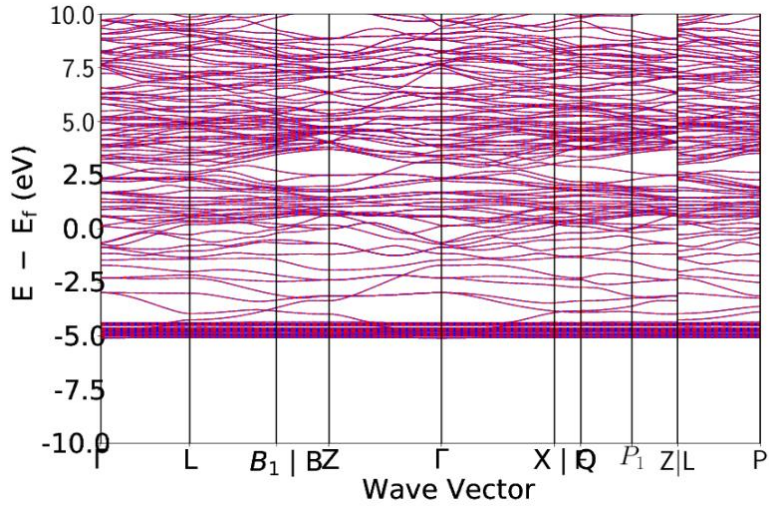 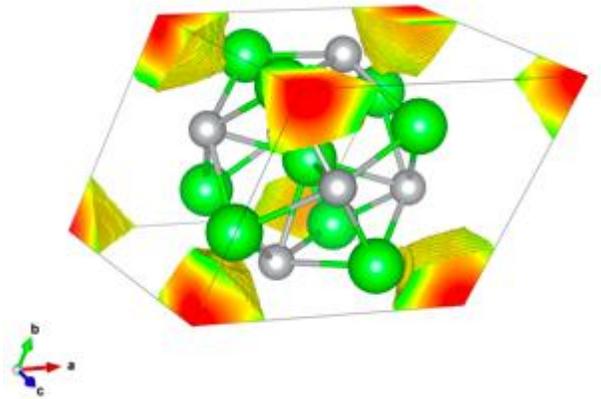

mp-30366; $Ca_3Au$. Symmetry: Orthorhombic; Pnma (number 62). Structure: $Ca_3Au$-type. Origin: ICSD;[39] comments: N/A. Anionic electron coordination environment: octahedral with Ca.

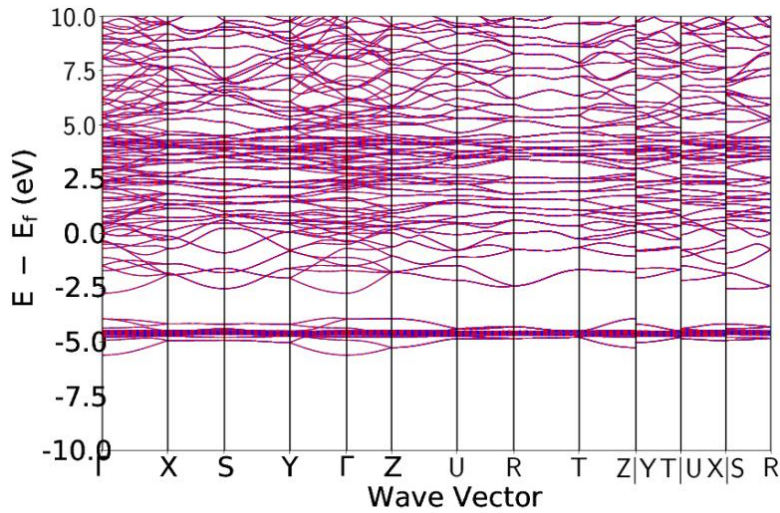 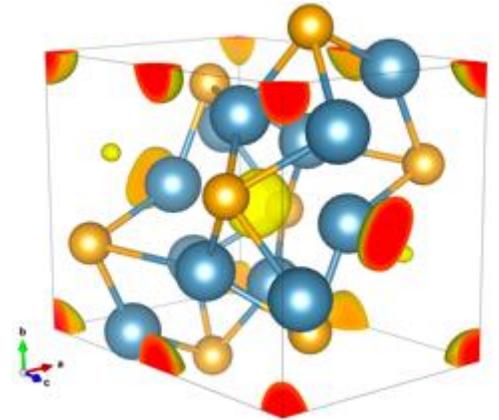

mp-30367; $Ca_5Au_2$. Symmetry: Monoclinic; C2/c (number 15). Structure: $Ca_5Au_2$-type. Origin: ICSD;[39] comments: N/A. Anionic electron coordination environment: octahedral with Ca.

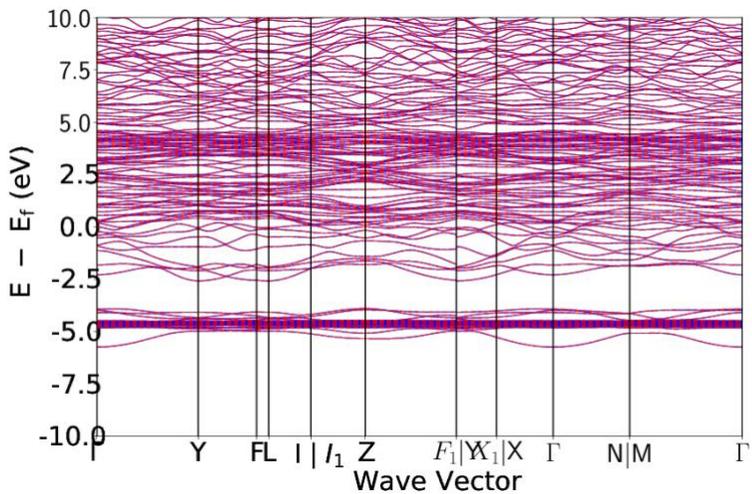 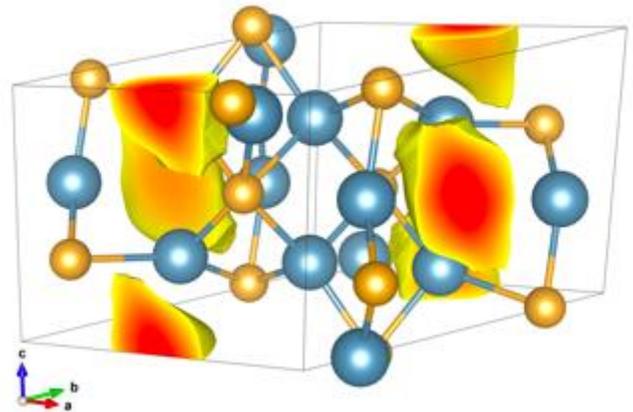

mp-30368; Ca$_5$Au$_3$. Symmetry: tetragonal; I4/mcm (number 140). Structure: Cr$_5$B$_3$-type. Origin: ICSD;[38] comments: N/A. Anionic electron coordination environment: tetrahedral with Ca.

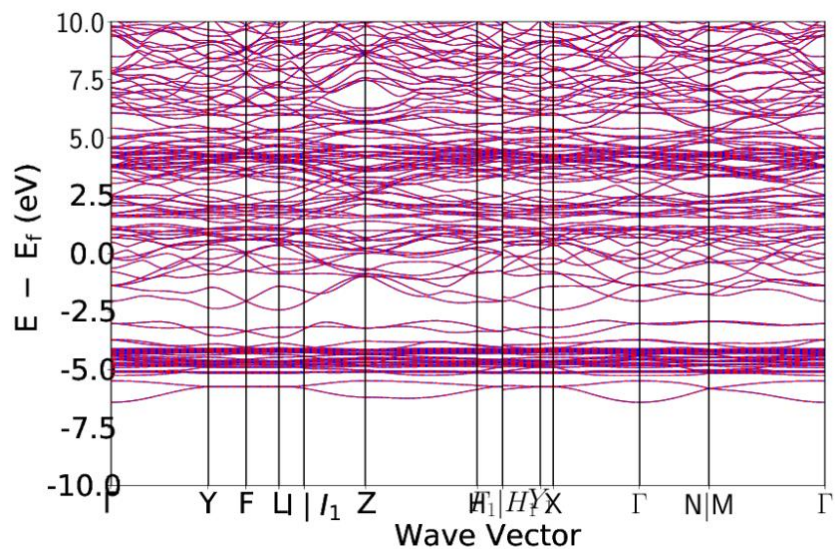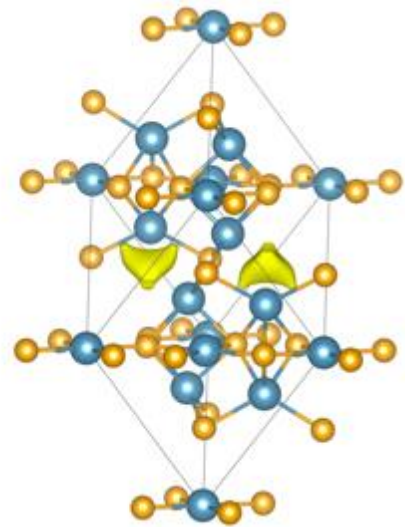

mp-30422; Sr$_7$Au$_3$. Symmetry: hexagonal; P6$_3$mc (number 186). Structure: Sr$_7$Au$_3$-type. Origin: ICSD;[38] comments: N/A. Anionic electron coordination environment: octahedral with Sr.

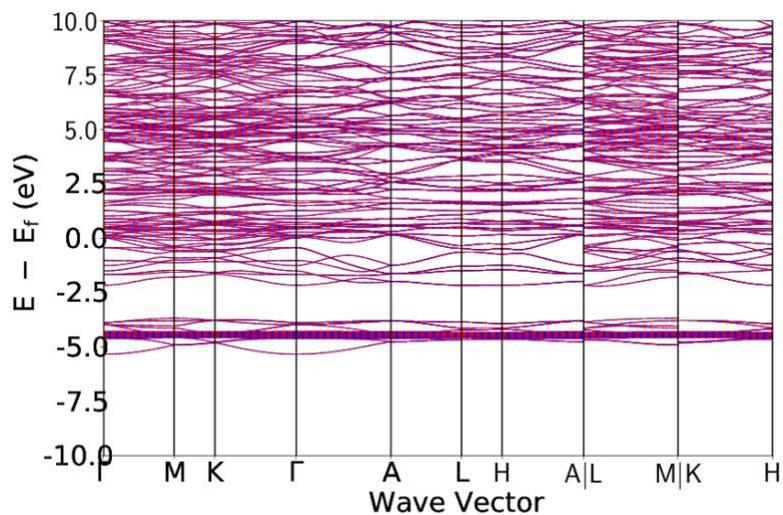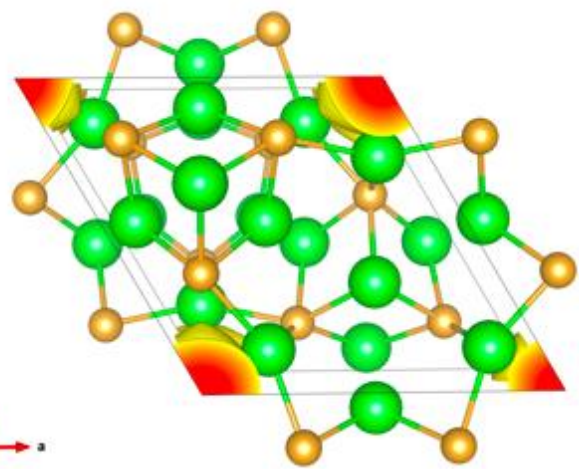

mp-371; La$_3$Tl. Symmetry: cubic; Pm-3m (number 221). Structure: Cu$_3$Au-type. Origin: ICSD;[29] comments: N/A. Anionic electron coordination environment: octahedral with La and Tl.

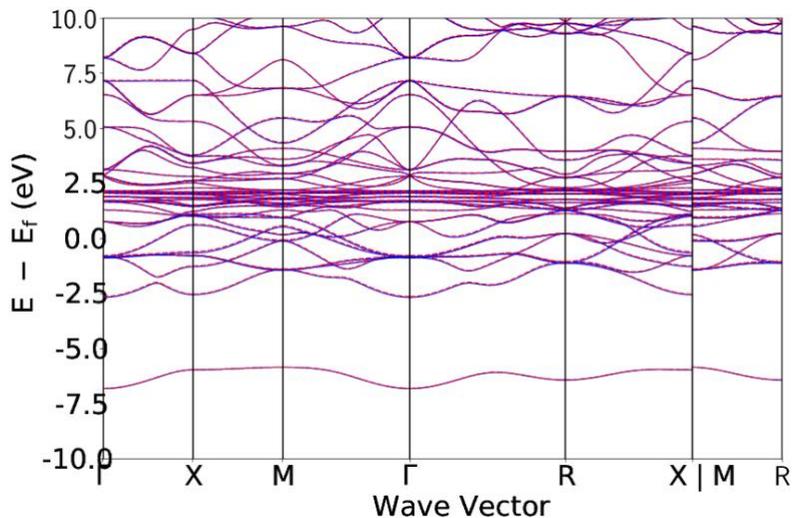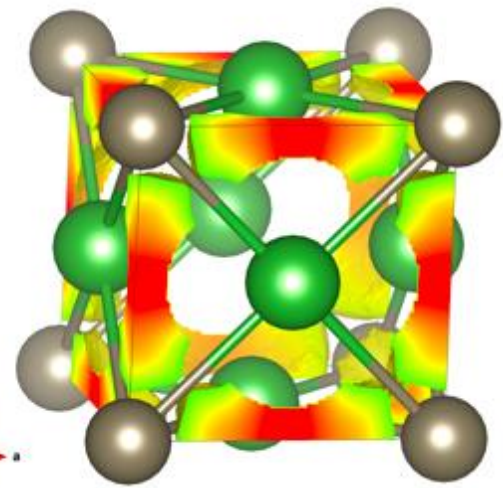

mp-4579; LaSiRu. Symmetry: tetragonal; P4/nmm (number 129). Structure: LaSiRu-type. Origin: ICSD;[40] comments: N/A. Anionic electron coordination environment: tetrahedral with La.

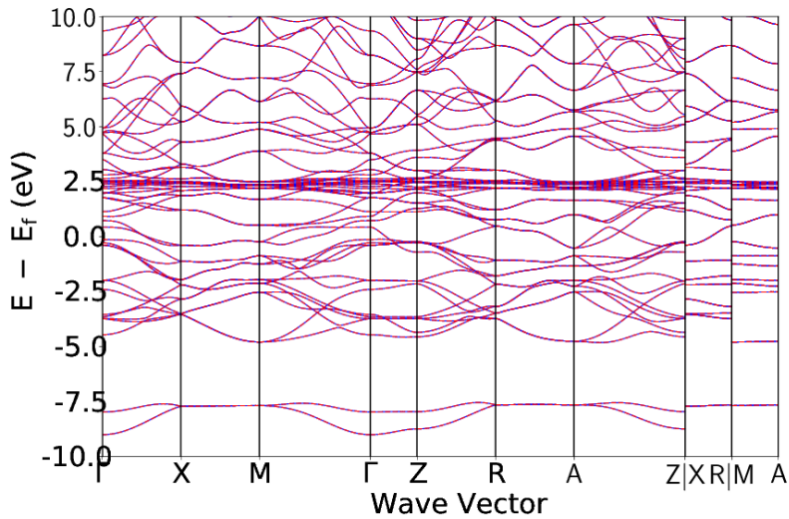 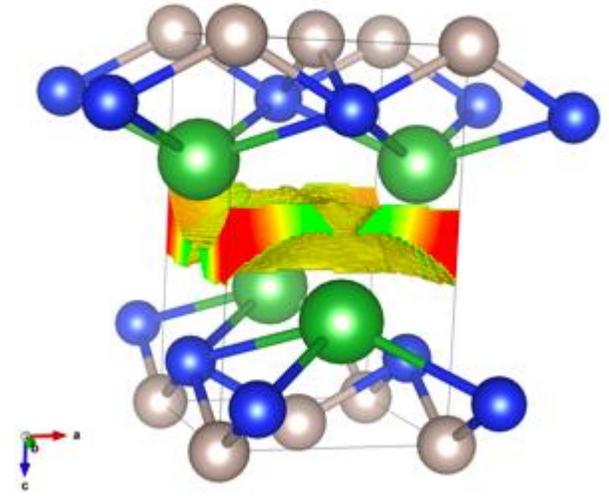

mp-4738; PrScGe. Symmetry: tetragonal; I4/mmm (number 139). Structure: CeScSi-type. Origin: ICSD;[41] comments: "PrScGe demonstrates antiferromagnetic transition below $T_N$= 140 K, but below 83 K trends to ferromagnetic-type ordering". Anionic electron coordination environment: tetrahedral with Pr.

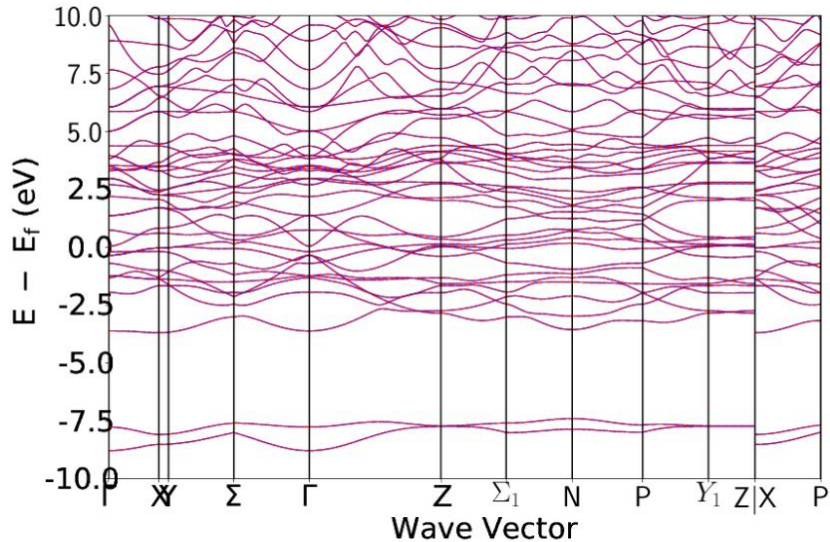 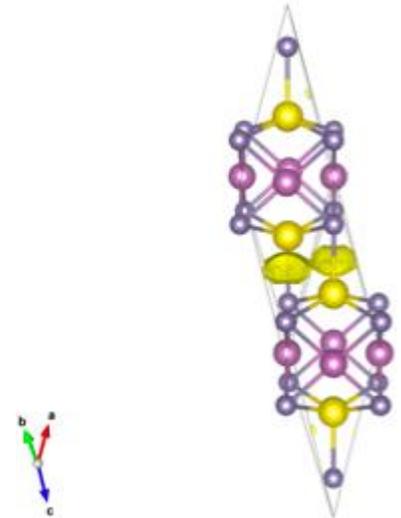

mp-4854; NdScGe. Symmetry: tetragonal; I4/mmm (number 139). Structure: CeScSi-type. Origin: ICSD;[41] comments: N/A. Anionic electron coordination environment: tetrahedral with Nd.

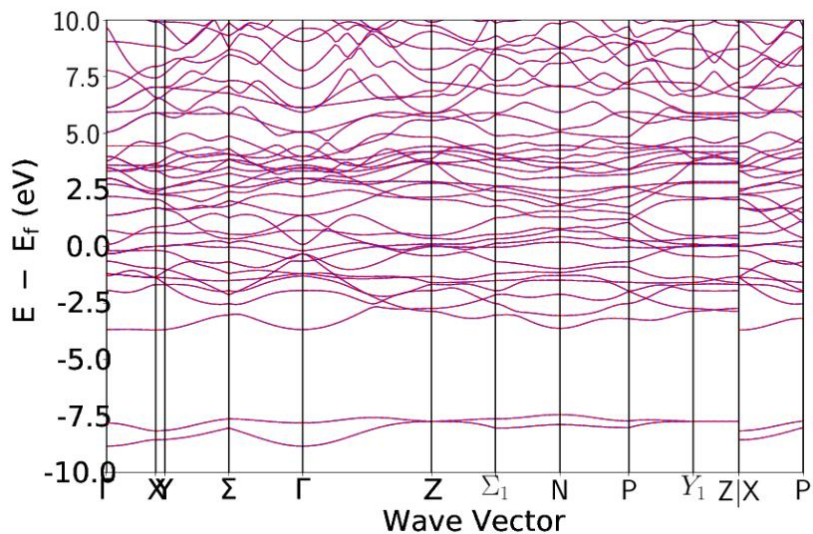 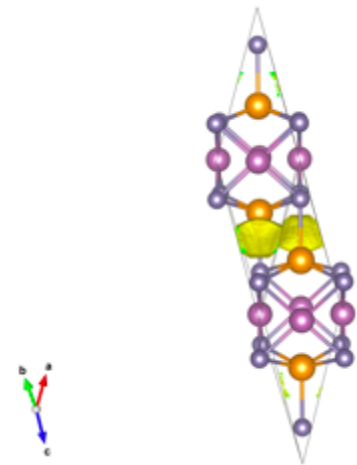

mp-542680; Na$_3$In$_2$Au. Symmetry: cubic; Fd-3m (number 227). Structure: NiTi$_2$-type. Origin: ICSD;[42] comments: compared to 'well-known Zintl-phase'. Anionic electron coordination environment: octahedral with Na.

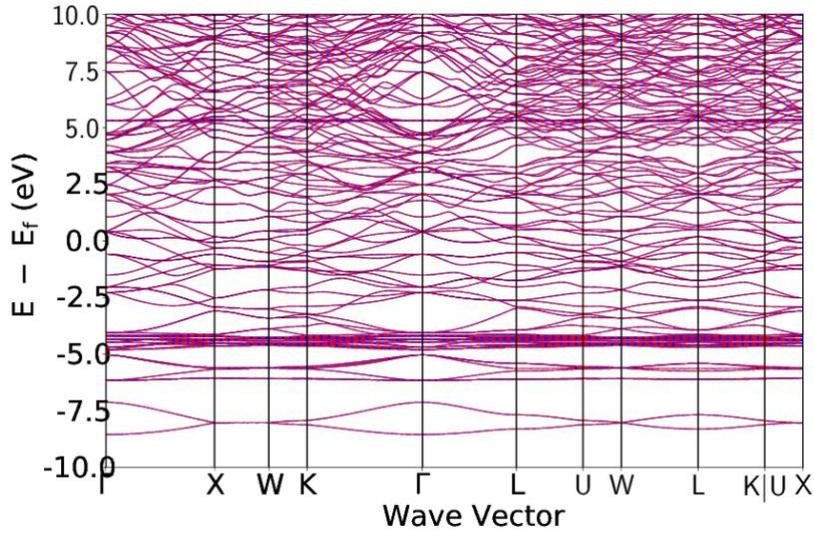 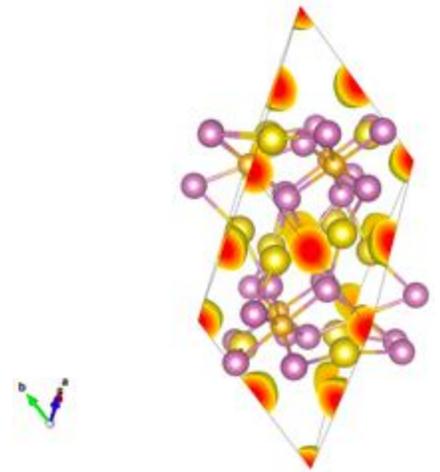

mp-542681; Na$_3$In$_2$Ag. Symmetry: cubic; Fd-3m (number 227). Structure: NiTi$_2$-type. Origin: ICSD;[42] comments: compared to 'well-known Zintl-phase'. Anionic electron coordination environment: octahedral with Na.

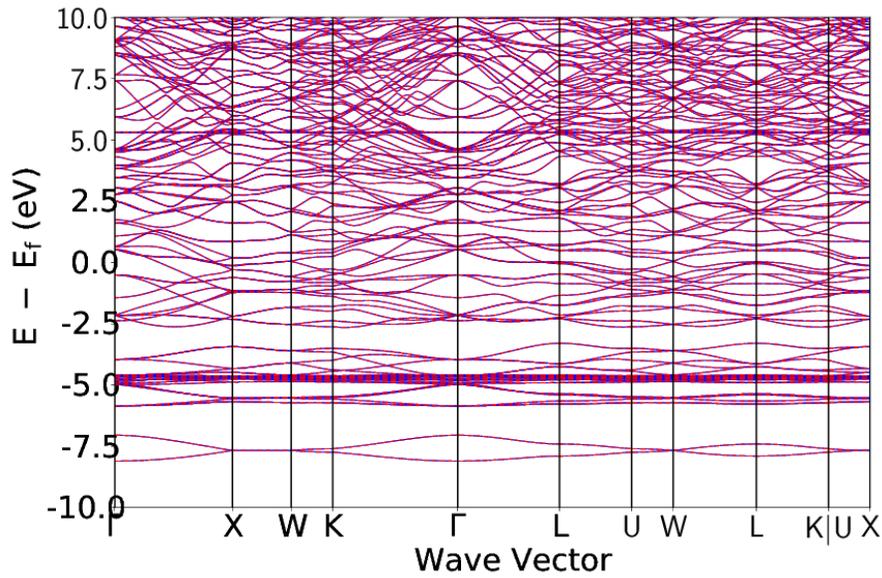 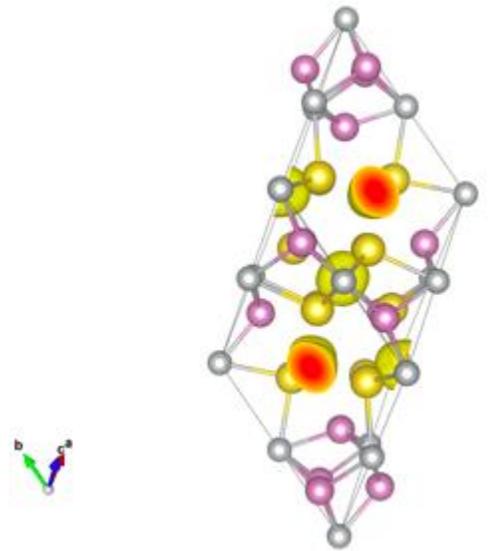

mp-567342; Nd$_4$MgIr. Symmetry: cubic; F-43m (number 216). Structure: NiTi$_2$-type. Origin: ICSD;[43] comments: referred to as intermetallic. Anionic electron coordination environment: octahedral with Nd.

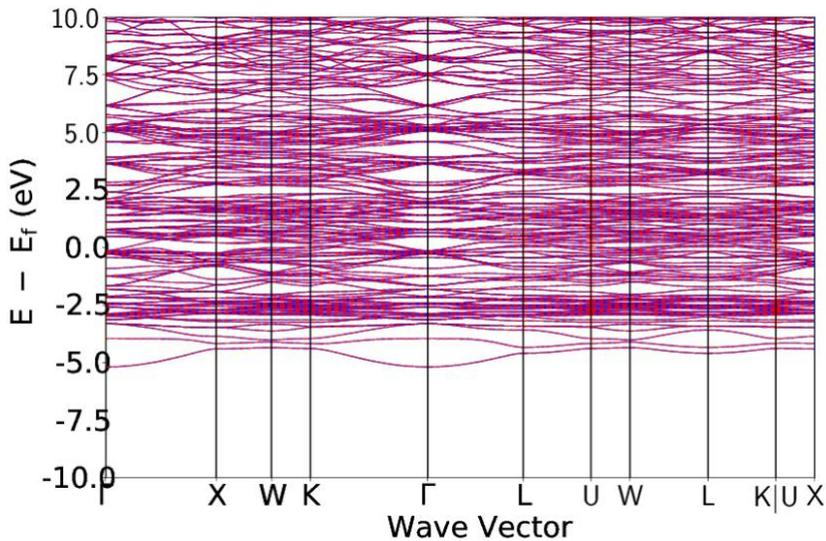 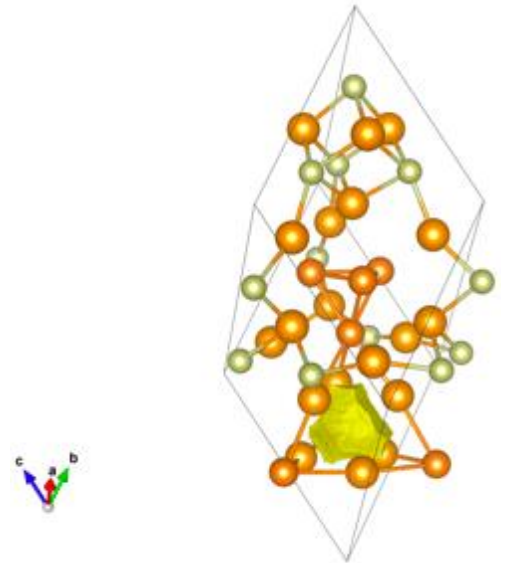

mp-569535; Ca$_2$Bi. Symmetry: tetragonal; I4/mmm (number 139). Structure: CeScSi-type. Origin: ICSD;[44] comments: N/A. Anionic electron coordination environment: octahedral with Ca.

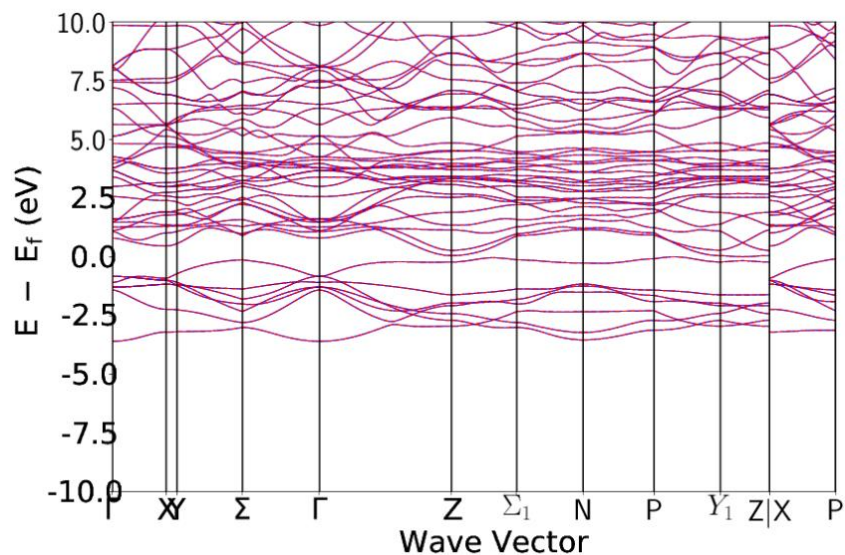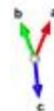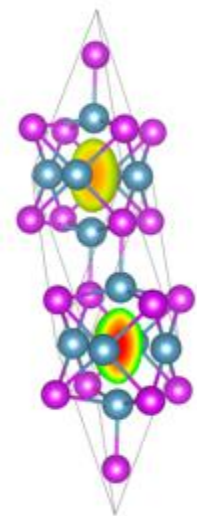

mp-570400; Ba$_7$Al$_{10}$. Symmetry: trigonal; R-3m (number 166). Structure: Ba$_7$Al$_{10}$-type. Origin: ICSD;[34] comments: referred to as Zintl compound. Anionic electron coordination environment: linear with Ba.

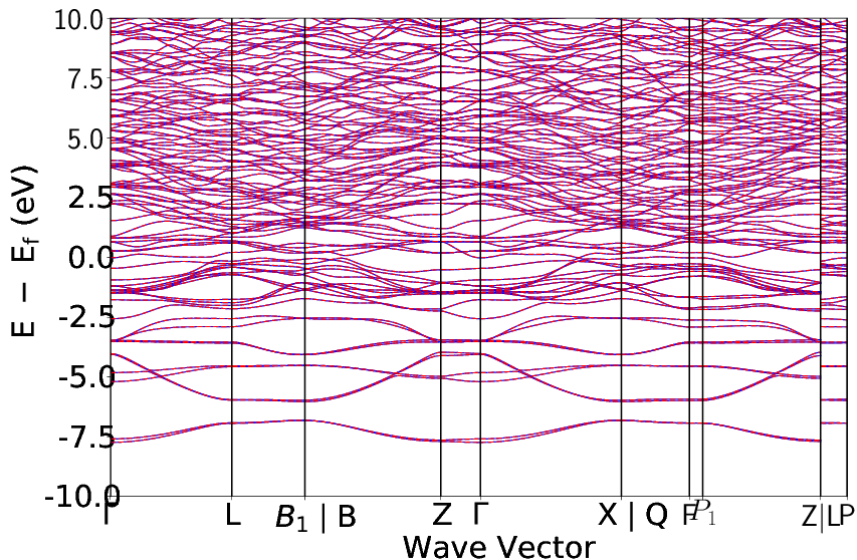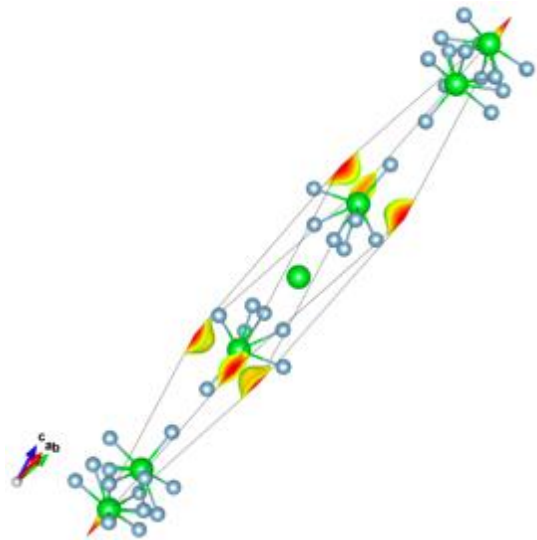

mp-573908; Sr$_{11}$(MgSi$_5$)$_2$. Symmetry: monoclinic; C2/m (number 12). Structure: Ba$_7$Al$_{10}$-type. Origin: ICSD;[45] comments: band splitting indicative of magnetism. Anionic electron coordination environment: linear with Sr.

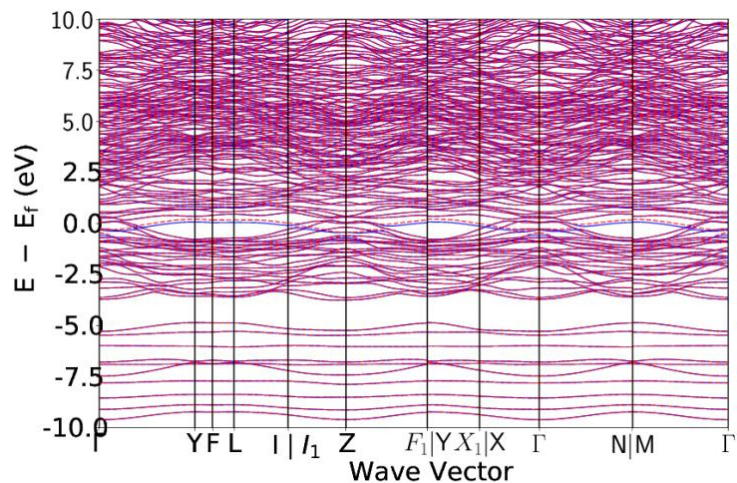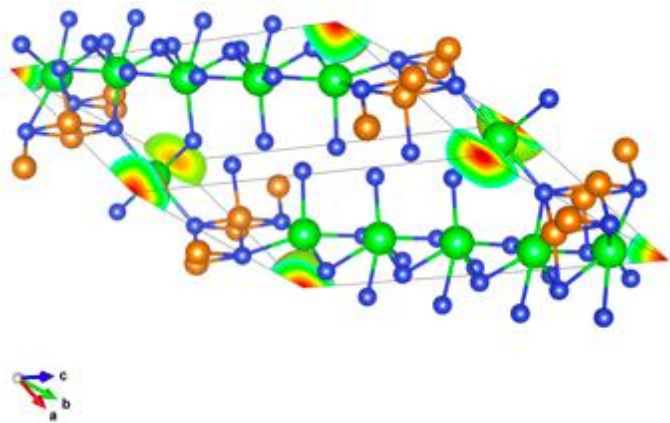

mp-605873; Pr$_4$MgRu. Symmetry: cubic; F-43m (number 216). Structure: NiTi$_2$-type. Origin: ICSD;[48] comments: N/A. Anionic electron coordination environment: trigonal prismatic with Pr.

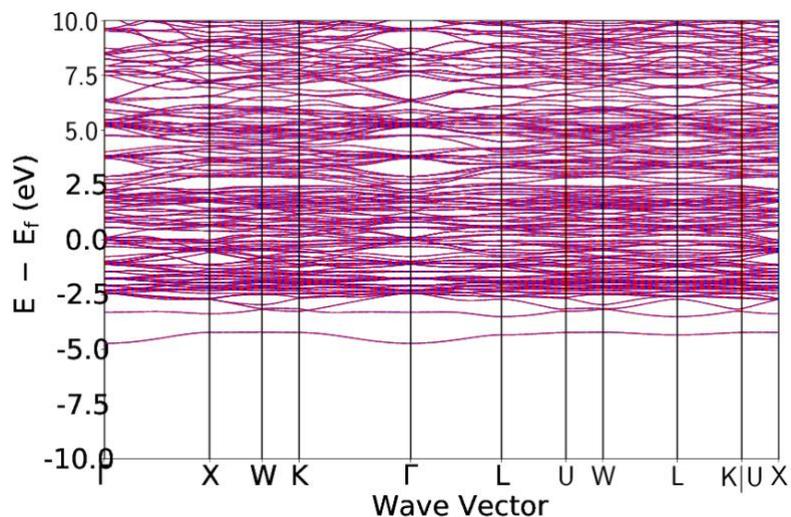

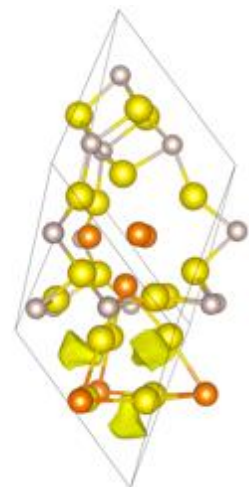

mp-630923; Ba$_3$Pb$_5$. Symmetry: orthorhombic; Cmcm (number 63). Structure: Ba$_3$Pb$_5$-type. Origin: ICSD;[49] comments: "Ba$_3$X$_5$ (X = Sn, Pb) and Sr$_3$Sn$_5$ fall into the region between the classical Zintl phases and intermetallic compounds". Anionic electron coordination environment: linear with Ba.

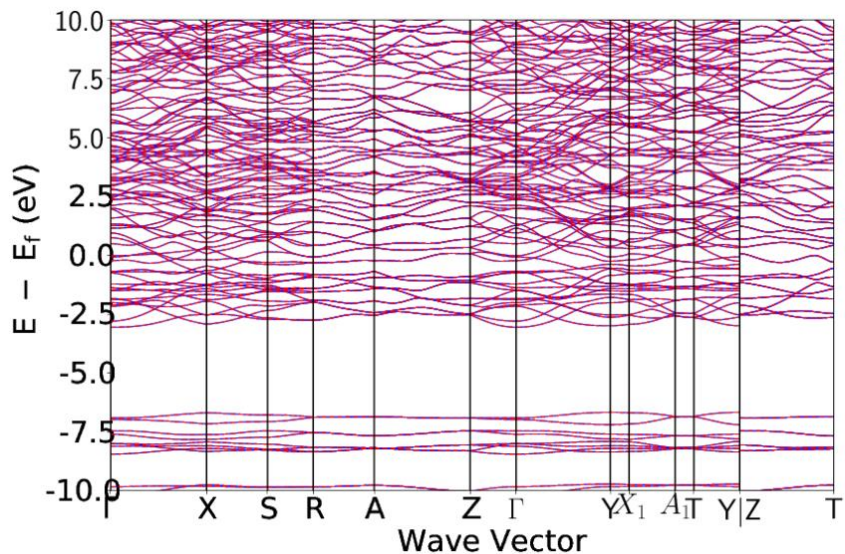

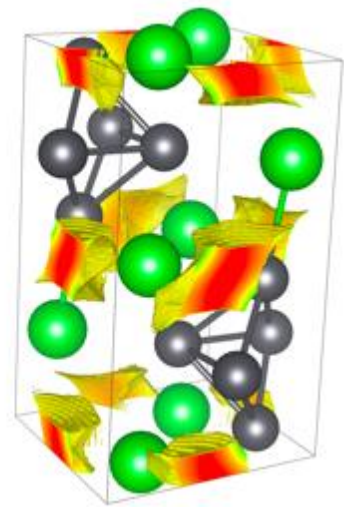

mp-645130; Pr$_4$MgCo. Symmetry: cubic; F-43m (number 216). Structure: NiTi$_2$-type. Origin: ICSD;[50] comments: referred to as intermetallic. Anionic electron coordination environment: octahedral with Pr.

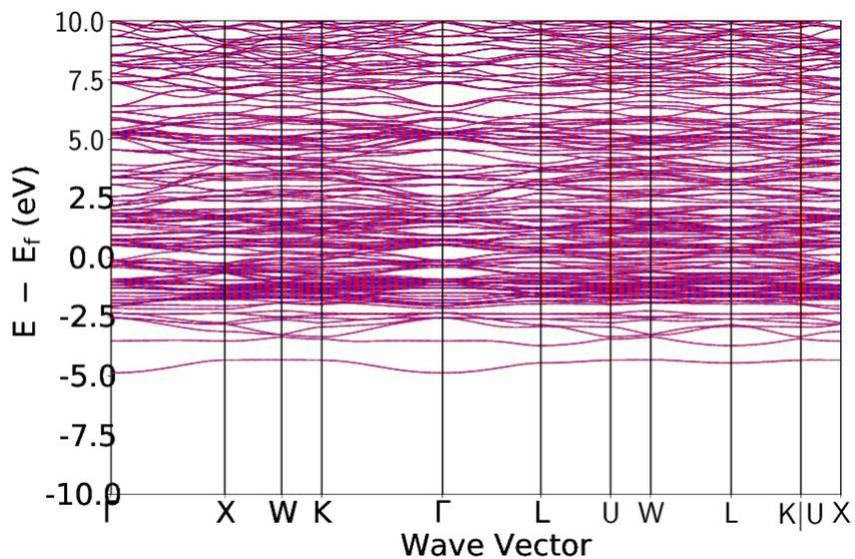

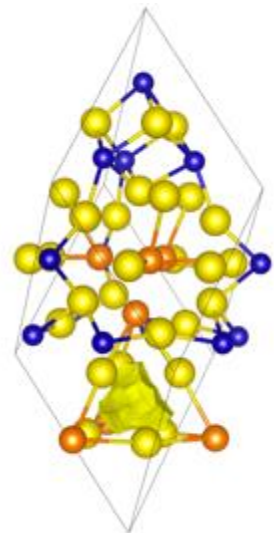

mp-693; Pr$_3$Tl. Symmetry: cubic; Pm-3m (number 221). Structure: Cu$_3$Au-type. Origin: ICSD;[51] comments: referred to as intermetallic. Anionic electron coordination environment: octahedral with Pr and Tl.

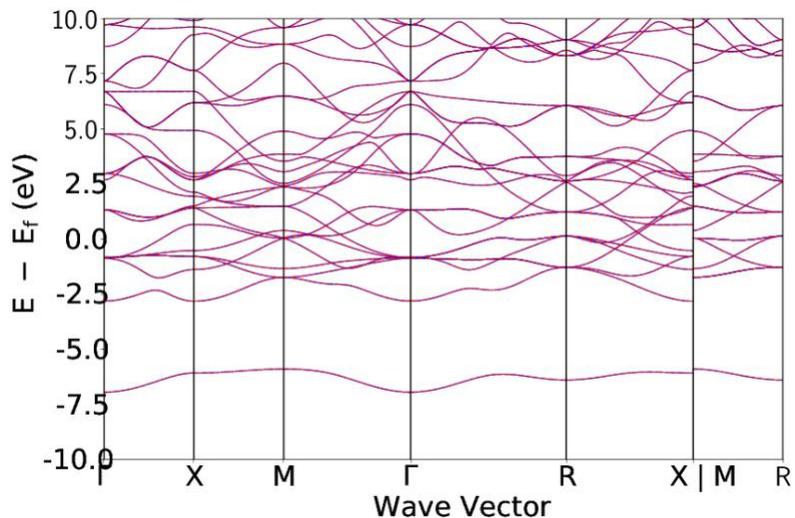
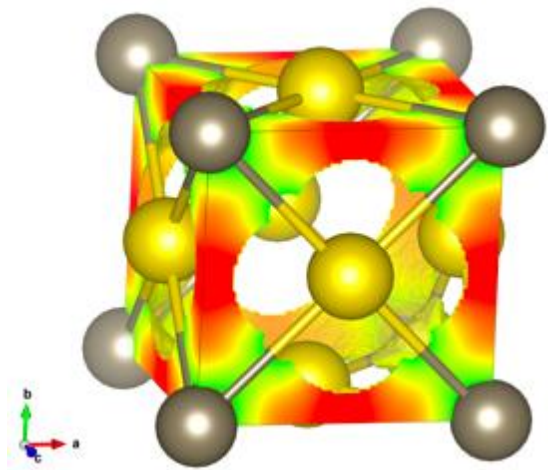

mp-7376; Sr$_3$(AlSn)$_2$. Symmetry: orthorhombic; Immm (number 71). Structure: Ta$_3$B$_4$-type. Origin: ICSD;[52] comments: N/A. Anionic electron coordination environment: distorted square pyramidal with Sr, Al and Sn.

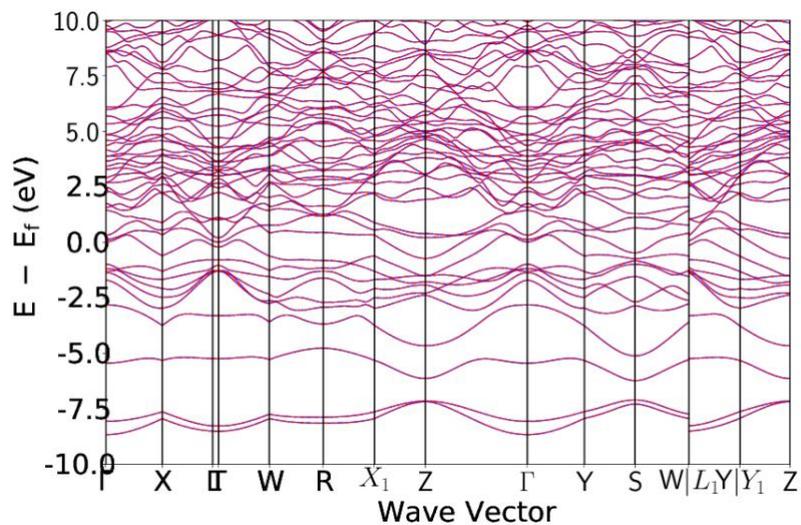
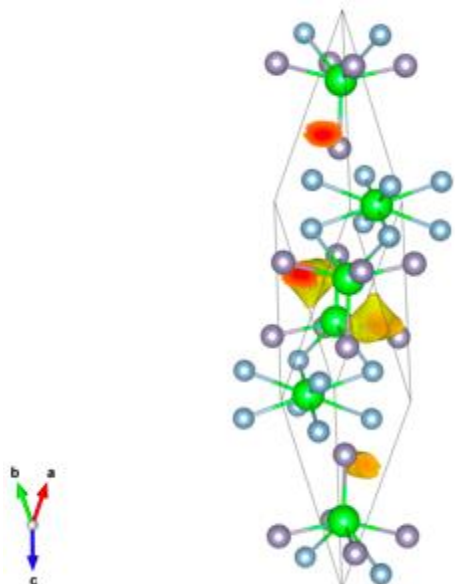

mp-746; Sr$_5$Si$_3$. Symmetry: tetragonal; I4/mcm (number 140). Structure: Cr$_5$B$_3$-type. Origin: ICSD;[13] comments: *ab inito* predicted structure. Anionic electron coordination environment: tetrahedral with Sr.

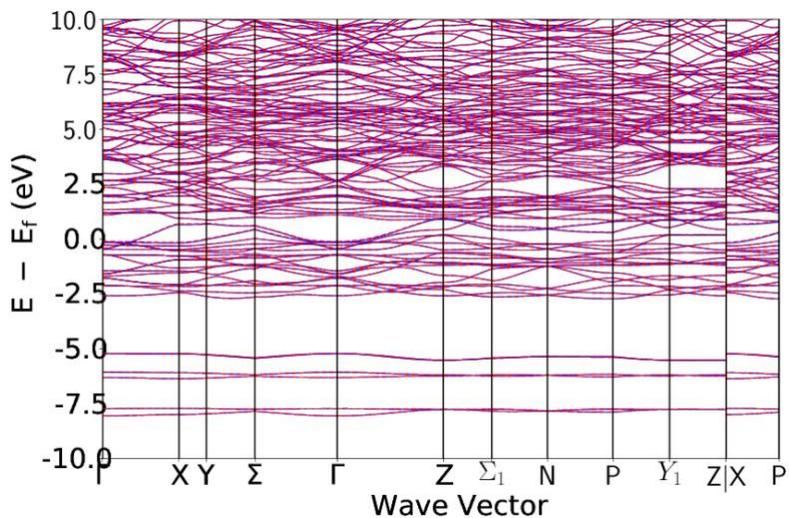
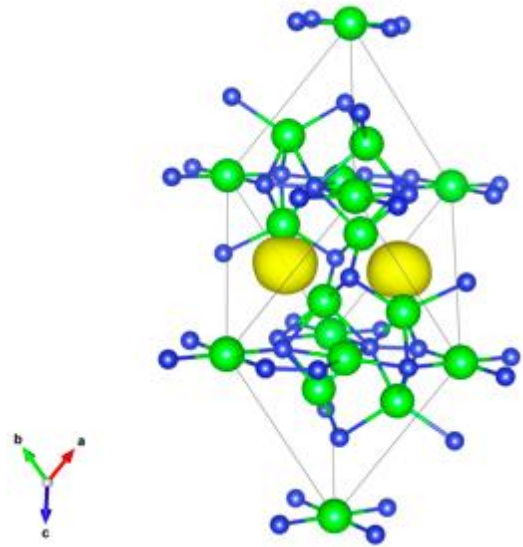

mp-7507; Sr$_3$Li$_2$. Symmetry: tetragonal; P4_2/mnm (number 136). Structure: Gd$_3$Al$_2$-type. Origin: ICSD;[53] comment: referred to as intermetallic. Anionic electron coordination environment: square pyramidal with Sr

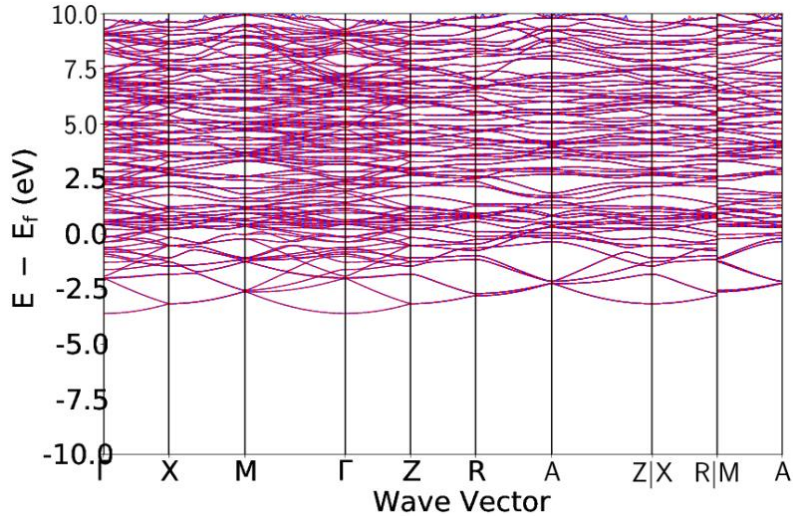
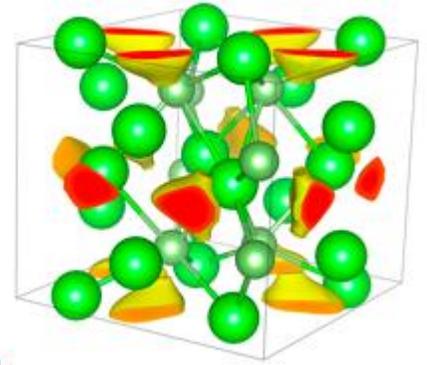

and Li.

mp-793; Ca$_5$Si$_3$. Symmetry: tetragonal; I4/mcm (number 140). Structure: Cr$_5$B$_3$-type. Origin: ICSD;[27] comments: referred go as Zintl phase. Suggest unresolved hydrogen might be included in structure. Anionic electron coordination environment: tetrahedral with Ca.

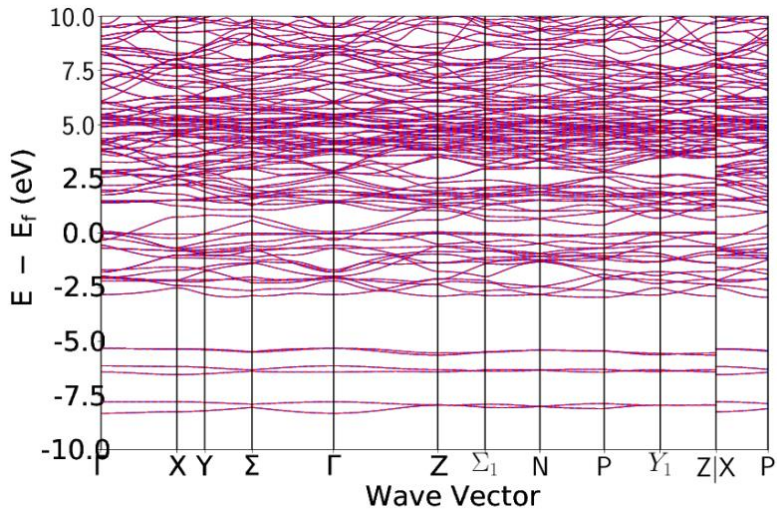
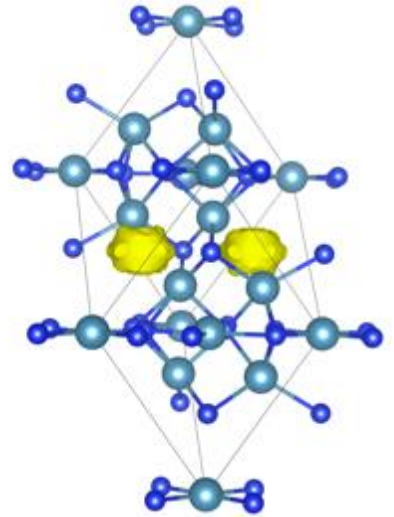

mp-8320; SmScSi. Symmetry: tetragonal; I4/mmm (number 139). Structure: CeScSi-type. Origin: ICSD;[54] comments: N/A. Anionic electron coordination environment: tetrahedral with Sm.

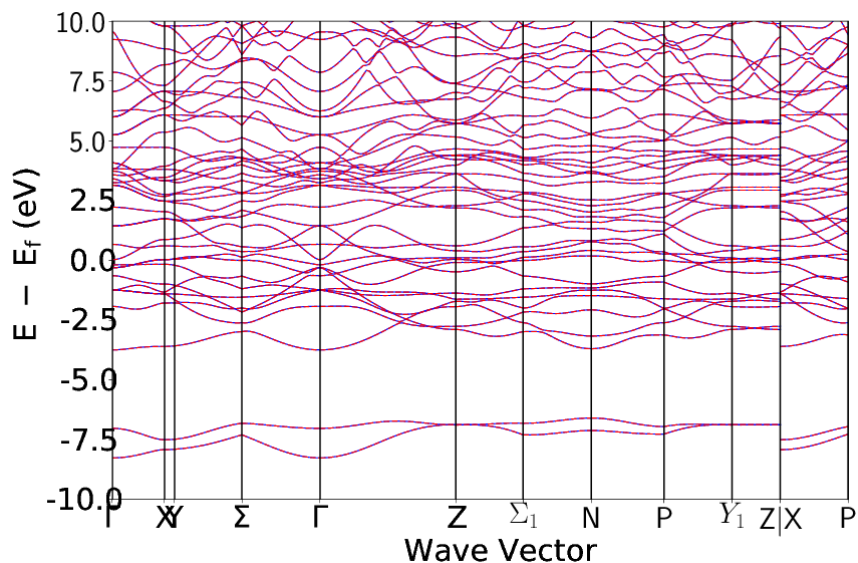
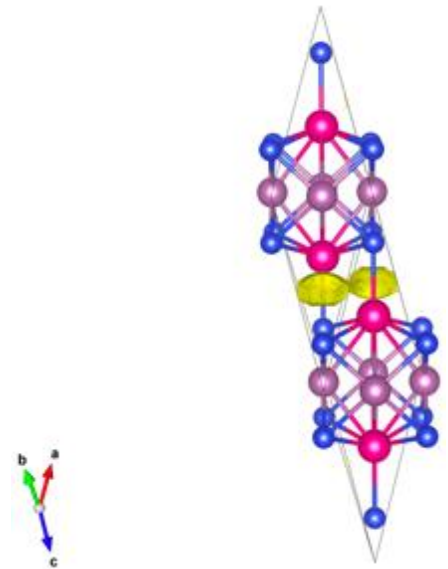

mp-872; BaSn. Symmetry: orthorhombic; Cmcm (number 63). Structure: CrB-type. Origin: ICSD;[22] comments: described as Zintl: "general valence equation suggests the presence of anion-anion chains", reacts violently with moisture. Anionic electron coordination environment: distorted octahedral with Ba and Sn.

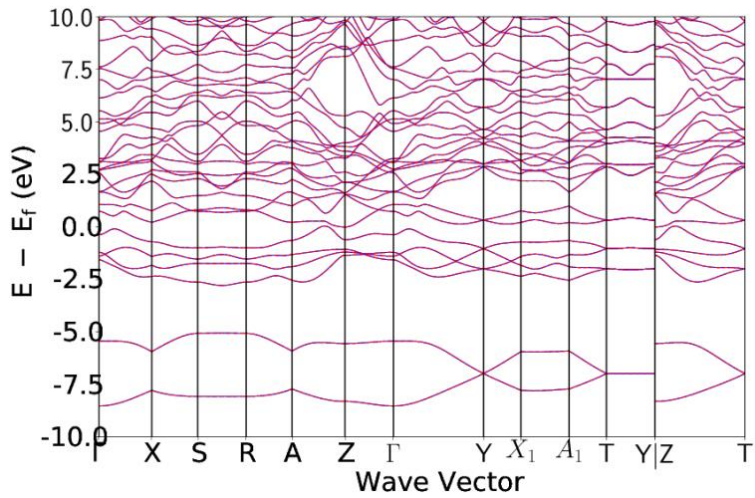 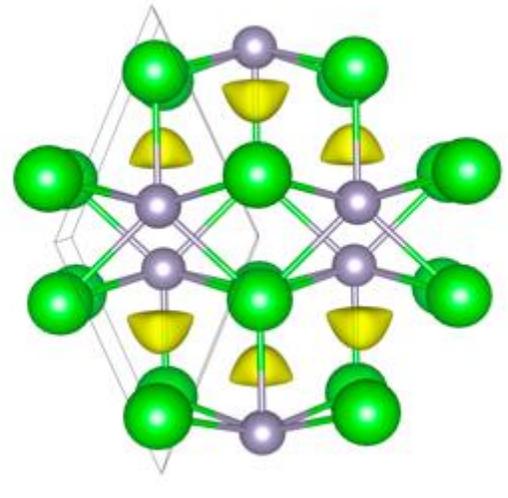

mp-973508; KRb$_3$. Symmetry: tetragonal; I4/mmm (number 221). Structure: KRb$_3$-type. Origin: OQMD;[55] comments: N/A. Anionic electron coordination environment: tetrahedral with K and Rb.

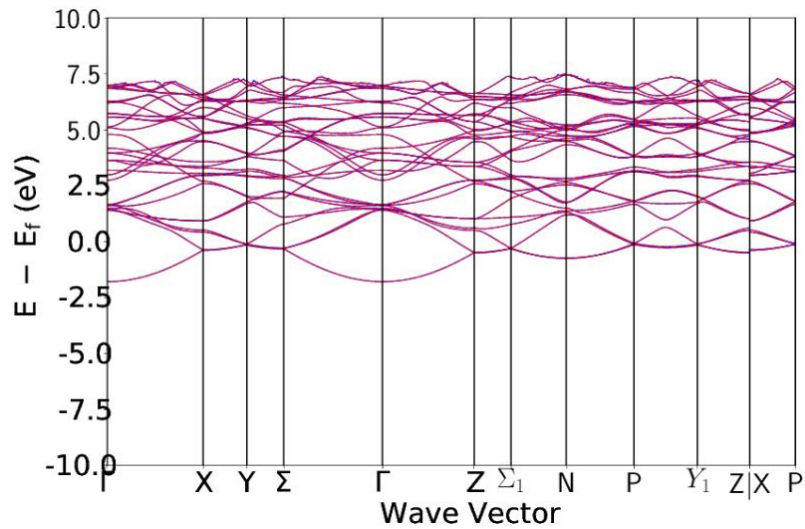 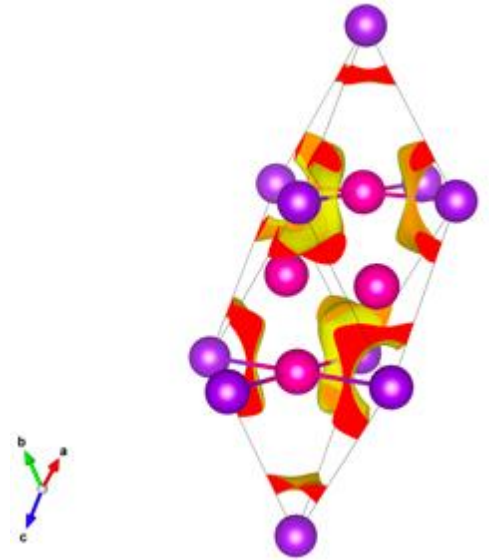

mp-974066; Nd$_3$Sm. Symmetry: cubic; Pm-3m (number 221). Structure: Cu$_3$Au-type. Origin: OQMD;[55] comments: N/A. Anionic electron coordination environment: octahedral with Nd and Sm.

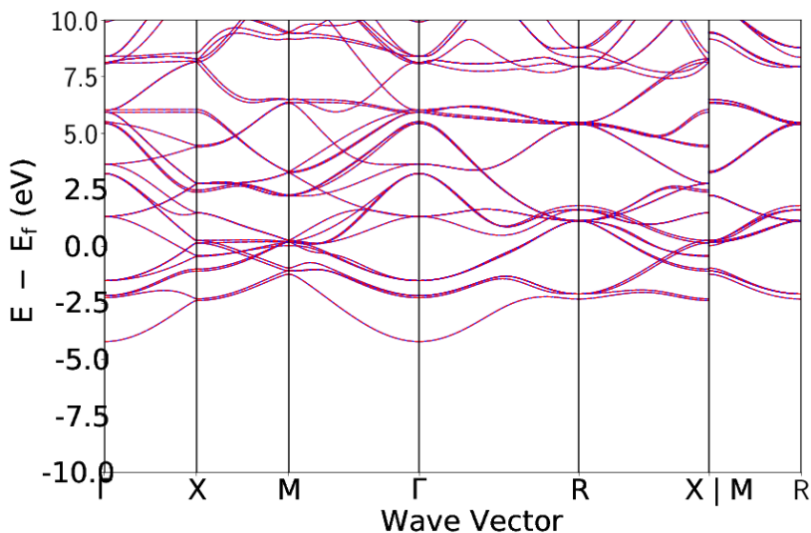 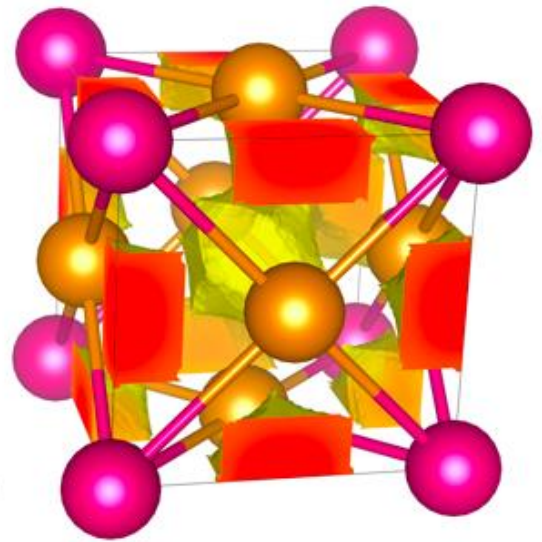

mp-976115; K$_3$Rb. Symmetry: hexagonal; P6_3/mmc (number 194). Structure: HCP-metal. Origin: OQMD;[55] comments: N/A. Anionic electron coordination environment: trigonal planar with K.

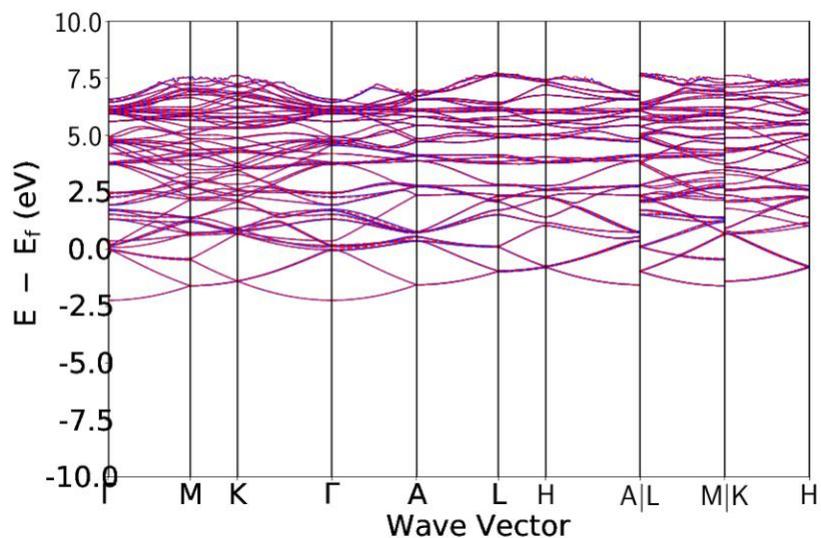
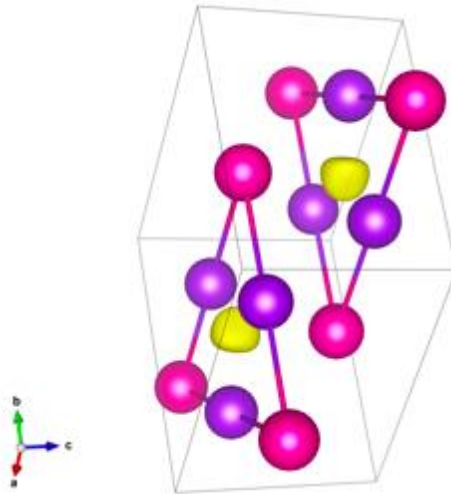

mp-984744; Ca$_3$Tl. Symmetry: hexagonal; P6_3/mmc (number 194). Structure: HCP-metal. Origin: OQMD;[55] comments: N/A. Anionic electron coordination environment: octahedral with Ca.

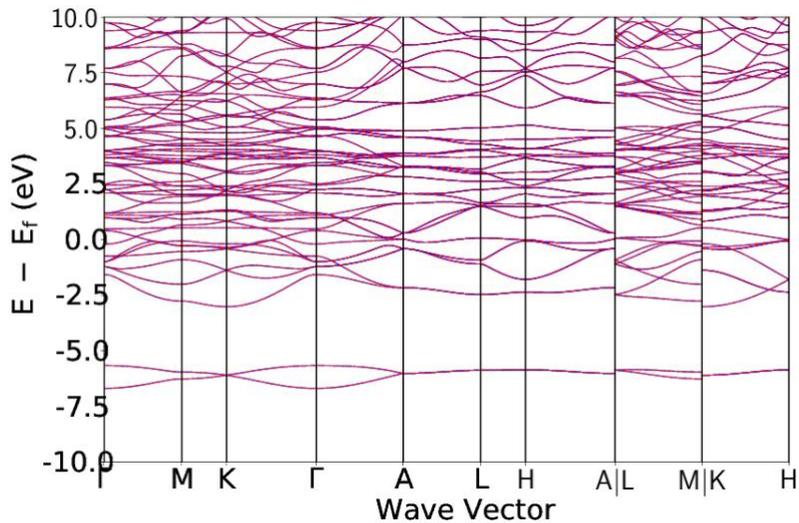
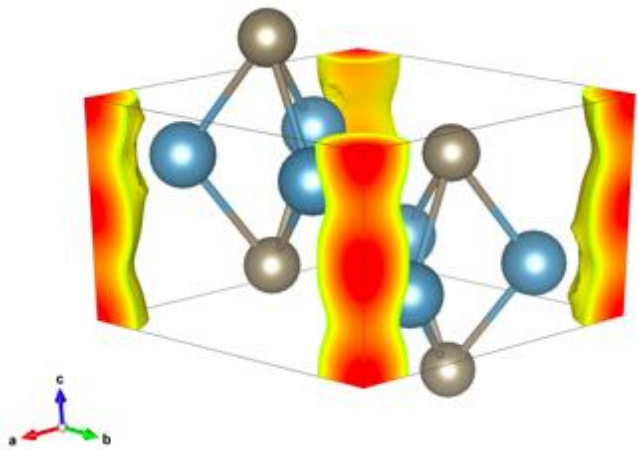

mp-9909; Ba$_5$Sb$_3$. Symmetry: hexagonal; P6_3/mcm (number 193). Structure: Mn$_5$Si$_3$-type. Origin: ICSD;[37] comments: N/A. Anionic electron coordination environment: octahedral with Ba.

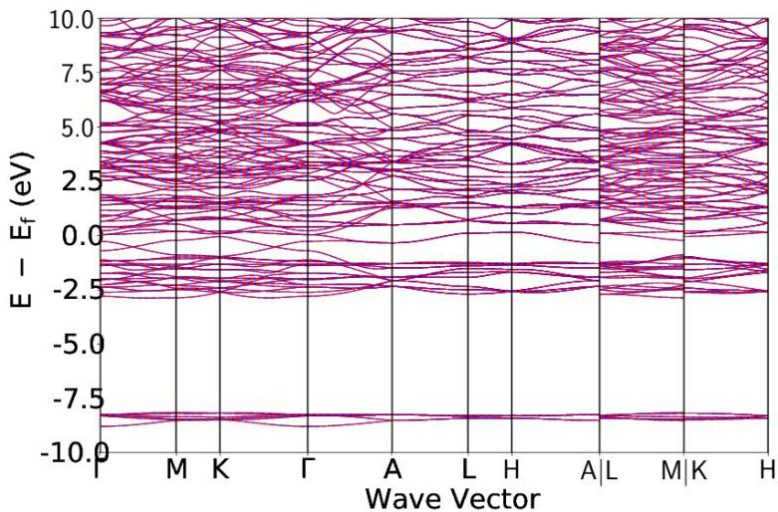
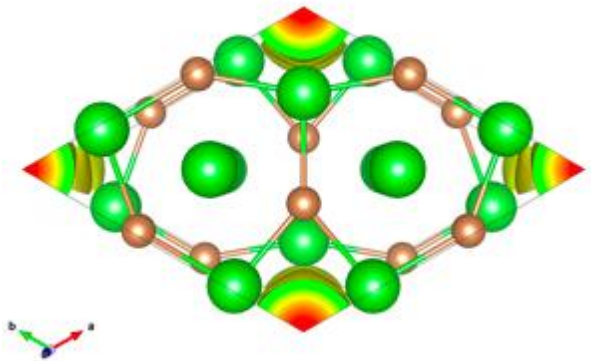